\documentclass[12pt]{article}

\usepackage{bbm}
\usepackage{color}
\usepackage{ulem}
\usepackage{url} 
\usepackage{latexsym,epsfig,amssymb,amsmath,slashed}

\usepackage{csquotes}

\newcommand{\be}{\begin{equation}}
\newcommand{\ee}{\end{equation}}
\newcommand{\bea}{\begin{eqnarray}}
\newcommand{\eea}{\end{eqnarray}}

\def\bar{\overline}
\def\tilde{\widetilde}
\def\sma{S_{\mu\alpha}}
\def\timn{T_{\mu\nu}}
\def\thetab{\bar\theta}
\def\nmn{\eta_{\mu\nu}}
\catcode`@=12
\usepackage[numbers,compress]{natbib}
\numberwithin{equation}{section}

\usepackage[bookmarks=true,colorlinks=true,linkcolor=black,citecolor=black,urlcolor=black]{hyperref} 

\begin{document}
\begin{titlepage}
\begin{flushright}
SISSA 48/2013/FISI\\
WIS/11/13-OCT-DPPA
\end{flushright}
\bigskip
\def\thefootnote{\fnsymbol{footnote}}

\begin{center}
\vskip -10pt
{\LARGE
{\bf
Supercurrent multiplet correlators \\ 
\vspace{0.25in}
at weak and strong coupling
}
}
\end{center}

\bigskip
\begin{center}
{\large
Riccardo Argurio,$^{1,2}$
Matteo Bertolini,$^{3,4}$
Lorenzo Di Pietro,$^{3,5}$ \\
\vskip 5pt
Flavio Porri$^3$ and
Diego Redigolo$^{1,2}$}

\end{center}

\renewcommand{\thefootnote}{\arabic{footnote}}

\begin{center}
$^1${Physique Th\'eorique et Math\'ematique \\
Universit\'e Libre de Bruxelles, C.P. 231, 1050 Bruxelles, Belgium\\}
$^2${International Solvay Institutes, Brussels, Belgium}\\
$^3$ {SISSA and INFN - Sezione di Trieste\\
Via Bonomea 265; I 34136 Trieste, Italy\\}
$^4$ {International Centre for Theoretical Physics (ICTP)\\
Strada Costiera 11; I 34014 Trieste, Italy\\}
$^5$ {Weizmann Institute of Science, Rehovot 76100, Israel}

\end{center}

\noindent
\begin{center} {\bf Abstract} \end{center}
Correlators of gauge invariant operators provide useful information on
the dynamics, phases and spectra of a quantum field theory. 
In this paper, we consider four dimensional $\mathcal N=1$ supersymmetric theories and focus our attention on the supercurrent multiplet. We give a complete
characterization of
two-point functions of operators belonging to such multiplet, like the
energy-momentum tensor and the supercurrent, and study the relations
between them. We discuss instances of weakly
coupled and strongly coupled theories, in which different symmetries,
like conformal invariance and supersymmetry, may be conserved and/or
spontaneously or explicitly broken.  For theories at strong coupling,
we exploit AdS/CFT techniques. We provide a holographic description of different
properties of a strongly coupled theory, including a 
realization of the Goldstino mode in a simple illustrative model.

\vfill

\end{titlepage}

\setcounter{footnote}{0}

\section{Introduction and motivations}

Two-point correlators of gauge invariant operators encode useful
data of a quantum field theory. 
For example, they
carry information on the dynamical phases of the theory, and on its
spectrum in given
superselection sectors, including both (massless or gapped) discrete 
and continuum bound states.

Given a class of theories in which
a certain operator is defined, it is often useful to parametrize its
correlators in terms of scalar form factors, after imposing Lorentz
invariance and the appropriate symmetry constraints. Such form factors
will then depend on the couplings that characterize the theory, and on
the vacuum. 
 
When a quantum field theory (QFT) is supersymmetric, operators are organized in supermultiplets. If the vacuum is supersymmetric, form factors of operators belonging to the same supermultiplet are related to each other. On the other hand, if supersymmetry is spontaneously broken, these relation will be valid only at high energies, and their violation at low scales can be seen as an effective probe of supersymmetry breaking. A concrete realization of this idea, and an example of its usefulness, is General Gauge Mediation \cite{Meade:2008wd}, where the supermultiplet at hand is that of a conserved current of the QFT. In that case, when the QFT is used as a hidden sector and  is coupled to the Standard Model by gauging the symmetry, the two-point functions fully encode the resulting soft masses.

Here, we will consider correlators of operators belonging to another multiplet, the supercurrent multiplet, which contains the stress-energy tensor and the supercurrent, i.e. the conserved current of supersymmetry, and as such is ubiquitous in a supersymmetric QFT. In addition, this multiplet contains an R-current, which, depending on the theory one is considering, gets identified with the superconformal R-current or some other R-current, which may or may not be conserved. We will provide a complete parametrization of the two-point functions of these operators in terms of form factors, and derive the supersymmetric relations among them. 
 
The universality of the supercurrent multiplet indicates that its correlators encode the very general features of a supersymmetric QFT. In particular, they are directly affected by the breaking of conformal invariance, R-symmetry and/or supersymmetry. For instance, when any of these symmetries is spontaneously broken, $1/k^2$ poles associated to the Goldstone modes appear in the relevant correlators. We will organize form factors in two distinct sets, one associated to the traceless part of the correlators, that computes the central charge at conformal fixed points, and another one which corresponds to the traces and is generated by the explicit breaking of conformal invariance.

Having in mind possible applications to hidden sectors in models of
gauge mediation, as well as conformal or nearly-conformal sectors of
beyond the Standard Model physics, or more generically confining
theories, one is often interested in theories at strong coupling. In
this case, ordinary field theory techniques are limited, and the
(possibly only) analytical tool one can use is holography, which
provides, in fact, a direct way to compute correlators of gauge
invariant operators at strong coupling
\cite{Gubser:1998bc,Witten:1998qj}. 

This  approach was used in \cite{Argurio:2012cd,Argurio:2012bi} in the context of holographic models 
of gauge mediation and in the present paper we pursue it further,  in view of wider applications. We will discuss a class of 
simple weakly coupled and strongly coupled
models, and compute supercurrent multiplet correlator form factors in
different dynamical regimes, using respectively ordinary field theory
techniques and AdS/CFT ones. The models we discuss are not only
interesting per s\'e but  also as useful playgrounds in view of
applications to richer set-ups, such as holographic
constructions within five-dimensional ${\mathcal N}=2$ gauged supergravity and, ultimately, supersymmetry breaking backgrounds in  
string theory. A thorough analysis of two-point
functions of the supercurrent multiplet may help, in fact, in discerning whether
supersymmetry breaking is explicit or spontaneous, and on the stability properties of the proposed backgrounds.

The rest of the paper is organized as follows. In section 2 we start recalling the structure of the supercurrent multiplet in four dimensions.  Depending on the theory at hand, this is more conveniently 
described by a Ferrara-Zumino (FZ) multiplet
\cite{Ferrara:1974pz} or a ${\mathcal R}$ multiplet \cite{Gates:1983nr}.\footnote{We will not consider situations in which none of the two 
supermultiplets can be defined, and one should resort to the so-called ${\mathcal S}$ multiplet \cite{Clark:1978jx,Komargodski:2010rb}. See \cite{Komargodski:2010rb} for a detailed discussion.}  Then, we give the parametrization of the two-point functions in terms of form factors, and we derive the constraints imposed by supersymmetry and conformal symmetry. In section 3 we compute supercurrent correlators in simple weakly coupled examples, enjoying different patterns of symmetry breaking. We consider models where (super)conformal invariance is preserved, spontaneously or explicitly broken, as well as models where supersymmetry is spontaneously broken. In section 4 we repeat the same program for theories at strong coupling, considering the simplest holographic set-up one can think of, namely a five dimensional hard wall background \cite{ArkaniHamed:2000ds,Rattazzi:2000hs,Polchinski:2001tt,BoschiFilho:2002vd,Erlich:2005qh}, and use holography to extract two-point functions. This will provide a holographic realization of a 
variety 
of different dynamical behaviors, including, e.g. a holographic description of the Goldstino mode. We end in section 5 with a summary of our results and an outlook.

\section{Supercurrent multiplets and two-point functions}
\label{formfactors}

In any supersymmetric field theory one can define an energy--momentum tensor $T_{\mu\nu}$ and a supercurrent $S_{\mu\alpha}$ (i.e.~the Noether's current associated to supersymmetry) which are both conserved on-shell and can be accommodated in a supercurrent multiplet, the most widely known being the Ferrara-Zumino (FZ) multiplet \cite{Ferrara:1974pz}. 
 
The FZ multiplet can be described\footnote{Here and in the following we adhere to the conventions of \cite{Wess:1992cp}.}
by a pair of superfields $(\mathcal J_\mu,\, X)$ satisfying the relation
\begin{align}
 -2\, &\bar D^{\dot\alpha}\sigma^\mu_{\alpha\dot\alpha}\, \mathcal J_\mu = D_\alpha X\,, \label{FZ}
\end{align}
with $ \mathcal J_\mu $ being a real superfield, $ \mathcal J_\mu =  \mathcal J_\mu^* $,  and $X$ a chiral superfield, $\bar D_{\dot\alpha} X = 0$.
From the defining equation above one can work out the component expression of these two superfields. They read
\begin{equation}
\begin{split}
 \mathcal J_\mu =& j_\mu + \theta\left( S_\mu - \frac 1 3 \sigma_\mu \bar S \right) + \bar \theta\left( \bar S_\mu + \frac 1 3 \bar\sigma_\mu S \right) +\frac i 2 \theta^2\partial_\mu x^* -\frac i 2 \bar\theta^2\partial_\mu x 
\\&+ \theta\sigma^\nu \thetab\left( 2\timn-\frac 2 3 \nmn T+\frac 1 2 \varepsilon_{\mu\nu\rho\sigma}\partial^\rho j^\sigma\right) +\dots \label{J-comp}
\end{split}
\end{equation}
and
\begin{equation}
 X = x + \frac 2 3 \theta S + \theta^2\left( \frac 2 3 T + i\,\partial^\mu j_\mu \right) + \dots\label{X-comp}
\end{equation}
where $\dots$ stand for the supersymmetric completion of the superfield and we have defined the trace operators $T\equiv T^\mu_{\phantom\mu\mu}$ and $S_\alpha \equiv \sigma^\mu_{\alpha \dot\alpha} \bar S_\mu^{\dot\alpha}$. All in all, the FZ multiplet contains a (in general non-conserved) current $j_\mu$, a symmetric and conserved $\timn$, a conserved $\sma$ and a complex scalar $x$. This makes a total of 12 bosonic $+$ 12 fermionic operators. 

From the above expression one can also see that whenever $X$ vanishes the current $j_\mu$ becomes conserved and all trace operators vanish. In this case the theory is superconformal and $j_\mu$ becomes the always present (and conserved) superconformal R-current. 

For theories with an R-symmetry, being it the superconformal one or any other, there exists an alternative supermultiplet accommodating the energy-momentum tensor and the supercurrent, the so-called $\mathcal R$ multiplet \cite{Gates:1983nr}. This is defined in terms of a pair of superfields $(\mathcal R_\mu,\, \chi_\alpha)$ satisfying
\begin{align}
 -2\, &\bar D^{\dot\alpha}\sigma^\mu_{\alpha\dot\alpha}\, \mathcal R_\mu = \chi_\alpha \,, \label{RM}
\end{align}
where $\mathcal R_\mu$ is a real superfield ,  $\mathcal R_\mu =  \mathcal R_\mu^*$, and $\chi_\alpha$ a chiral superfield, $\bar D_{\dot\alpha} \chi_\alpha = 0$ which satisfies the identity $\bar D_{\dot\alpha} \bar\chi^{\dot\alpha} - D^\alpha \chi_\alpha=0$; this implies, in turn, that $\partial^\mu \mathcal R_\mu = 0$. From the latter relation it follows that the lowest component of $\mathcal R_\mu$ is indeed a conserved current. 
The component expression of the superfields making-up the $\mathcal R$ multiplet reads
\begin{equation}
\mathcal R_\mu = j_\mu + \theta S_\mu + \bar\theta \,\bar S_\mu + \theta\sigma^\nu \thetab\left( 2\timn + \frac 1 2 \varepsilon_{\mu\nu\rho\sigma} (\partial^\rho j^\sigma + C^{\rho\sigma}) \right) +\dots \label{R-comp}
\end{equation}
and 
\begin{equation}
\chi_\alpha =  - 2  S_\alpha - \left( 4 \delta_\alpha^\beta T + 2 i \,(\sigma^\rho \bar \sigma^\tau)_\alpha^\beta C_{\rho\tau}\right) \theta_\beta + 2 i\, \theta^2 \sigma^\nu_{\alpha\dot\alpha} \partial_\nu \bar S ^{\dot\alpha} +\dots \label{chi-comp}
\end{equation}
where again  $\dots$ stand for the supersymmetric completion, while
$C_{\mu\nu}$ is a closed two-form. The number of on-shell degrees of
freedom is 12 bosonic + 12 fermionic, as for the FZ multiplet. In a
theory where both the FZ and the ${\mathcal R}$ multiplets can be defined,
they are related by a shift transformation \cite{Komargodski:2010rb} (which acts as an improvement
 on $T_{\mu\nu}$ and $S_{\mu\alpha}$) defined as 
\begin{equation}
\label{shift1}
{\mathcal R}_\mu = {\mathcal J}_\mu + \frac 14 \bar\sigma_\mu^{\dot\alpha \alpha} \left [D_\alpha ,\,\bar D_{\dot\alpha} \right] U\quad,\quad X= -\frac 12 \bar D^2 U\quad,\quad\chi_\alpha= \frac 32 \bar D^2D_\alpha U\,, 
\end{equation}
where $U$ is a real superfield.

While in this paper we will not be concerned with theories where the
FZ multiplet cannot be defined
\cite{Komargodski:2010rb,Komargodski:2009pc}, it can sometime be
interesting, provided an R-symmetry is present, to consider the
$\mathcal R$ multiplet, instead. Such a situation typically occurs in
phenomenological models \cite{Nelson:1993nf}. For this reason, we will also discuss ${\mathcal R}$ multiplet correlators.

\subsection{Parametrization of two-point functions}
Let us start focusing on two-point functions of operators belonging to the FZ multiplet. One can use Poincar\'e invariance and conservation laws to fix completely the tensor structure of such correlators, and be left with a set of (model dependent) form factors. 

In euclidean momentum-space, the real correlators can be parametrized as follows
\begin{subequations}
\begin{align}
&\langle T_{\mu\nu}(k)\,T_{\rho\sigma}(-k)\rangle = -\frac 1 8 X_{\mu\nu\rho\sigma}\, C_2(k^2) -\frac 1 8 \frac{m^2}{k^2}\,(P_{\mu\nu}P_{\rho\sigma}-P_{\rho(\mu}P_{\nu)\sigma})\,F_2 (k^2)  \label{<TT>} \\
&\langle S_{\mu\alpha}(k)\, \bar{S}_{\nu\dot\beta}(-k)\rangle = - (Y_{\mu\nu})_{\alpha\dot\beta}\,C_{3/2}(k^2) -\frac{i}{2}\,m^2\,\varepsilon_{\mu\nu\rho\lambda}\,k^\rho\sigma^\lambda_{\alpha\dot\beta} \,F_{3/2}(k^2) + M^4 (\sigma_\mu \bar\sigma^\rho  \sigma_\nu)_{\alpha\dot\beta}
\frac{2 k_\rho}{k^2} 
\label{<SSbar2>} \\
&\langle j_\mu(k)\,j_\nu(-k)\rangle = -P_{\mu\nu}\, C_{1R}(k^2) - \frac{1}{3}\,m^2\eta_{\mu\nu}\,F_1(k^2)\label{<jj>} \\
&\langle x(k)\,x^*(-k)\rangle = \frac{2}{3}\,m^2 \,F_0(k^2) \label{<xxdagger>} \\
&\langle j_\rho(k)\,T_{\mu\nu}(-k) \rangle = i\,k_\rho\,P_{\mu\nu}\,I_3(k^2) \, \label{<jT>}
\end{align}
\end{subequations}
where $P_{\mu\nu}\equiv k^2 \eta_{\mu\nu} - k_\mu k_\nu$  is the transverse projector, and we have defined the traceless tensor 
\be
\label{X}
X_{\mu\nu\rho\sigma}=P_{\mu\nu}P_{\rho\sigma}-3P_{\rho(\mu}P_{\nu)\sigma}\,,
\ee
and its fermionic analog (by trace of the supercurrent operator we mean the contraction with $\sigma^\mu$)
\be
\label{Y}
(Y_{\mu\nu})_{\alpha\dot\beta}= k_\rho\sigma^\rho_{\alpha\dot\beta}\,P_{\mu\nu}+\frac{i}{2}\,k^2\,\varepsilon_{\mu\nu\rho\lambda}\,k^\rho\sigma^\lambda_{\alpha\dot\beta} \, .
\ee
In some terms a mass scale $m$ appears, which, as we will show below, is related to the explicit breaking of conformal invariance. Finally, a $1/k^2$ pole appears in the supercurrent correlator when supersymmetry is spontaneously broken at some scale $M$, defined by $\langle T_{\mu\nu}\rangle =  -M^4 \,\eta_{\mu\nu}$, signalling the presence of a Goldstino mode. Indeed, whenever supersymmetry is spontaneously broken, we have the (non-transverse) Ward identity 
 \begin{align}
 \langle(\partial^\mu S_{\mu\alpha})(k) \bar{S}_{\nu\dot{\beta}}(-k) \rangle &=  -\langle \delta_\alpha \bar{S}_{\nu\dot{\beta}} \rangle \,,\label{wimod}
 \end{align}
 where\footnote{The additional factor of $i$ with respect to the tranformations in appendix \ref{AppTrans} arises when the correlators are continued in Euclidean space.}
\begin{align}
\langle \delta_\alpha {\bar{S}}_{\mu\dot{\beta}} \rangle =  \langle \delta_{\dot{\beta}} S_{\mu\alpha}  \rangle = i {\sigma^\nu}_{\alpha \dot{\beta}} \, \langle 2 \, T_{\mu \nu} \rangle  \not = 0\,,
\end{align}
By substituting the parametrization (\ref{<SSbar2>}) of the
supercurrent two-point function in \eqref{wimod}, one easily sees that the above term provides the $1/k^2$ pole contribution.  

When appropriate, we have separated the structure of correlators in terms of a traceless and a trace part. The former is given by the functions $C_2$, $C_{3/2}$ and $C_{1R}$. Note that $C_2$ determines the central charge $c$ at a conformal fixed point. The form factors $F_2$, $F_{3/2}$, $F_1$, $F_0$ contribute instead to the trace operator correlators
\begin{subequations}\label{non-conformal}
\begin{align}
&\langle T(k)\,T(-k)\rangle = -\frac{3}{4} m^2 k^2F_2(k^2) \label{<TTtr>} \\
&\langle \bar S_{\dot\alpha}(k)\,  S_\alpha(-k)\rangle =  3 \sigma^\nu_{\alpha\dot\alpha} k_{\nu}\, m^2F_{3/2}(k^2) + 32 M^4 \frac{ \sigma^\nu_{\alpha\dot\alpha} k_{\nu}}{k^2}\label{<SSbartr>}\\
&k^{\mu} k^\nu\langle j_\mu(k)\,j_\nu(-k)\rangle=-\frac{1}{3}\,m^2 k^2\,F_1(k^2)\\
&\langle x(k)\,x^*(-k)\rangle = \frac{2}{3}\,m^2 \,F_0(k^2)\label{xxdagger2}\,.
\end{align}
\end{subequations}
Additional non-trivial two-point functions, given in terms of complex form factors, are
\begin{subequations}
\begin{align}
&\langle S_{\mu\alpha}(k)\, S_{\nu\beta}(-k) \rangle = m\,\varepsilon_{\alpha\beta}\,P_{\mu\nu}\,G_{3/2}(k^2) - 2i\,m\,\varepsilon_{\mu\nu\rho\lambda}\,k^\rho\,\sigma^{\lambda\tau}_{\alpha\beta}\, k_\tau\,\tilde G_{3/2}(k^2) \label{<SS>}\\
&\langle x(k)^*\,j_\mu(-k) \rangle = m\,k_\mu\,H_1(k^2) \label{<xj>} \\
&\langle x(k)^*\,T_{\mu\nu}(-k) \rangle = \frac{1}{2}\, m \,P_{\mu\nu}\,H_2(k^2)\,. \label{<xT>}
\end{align}
\end{subequations}
All in all, two-point functions can be parametrized in terms of eight real and four complex form factors.

\subsection{Supersymmetric relations among form factors}
On a supersymmetry preserving vacuum, the supersymmetry algebra imposes the following relations among form factors
\begin{subequations}\label{SUSYrel}
\begin{align}
\label{SUSYcomp}
 & C_2=C_{3/2}=C_{1R}\,,\quad F_2=F_{3/2}=F_1=F_0\, ,\quad I_3 = 0\,,
\\ 
\label{SUSYcomp2}
& H_2 = H_1 = G_{3/2} = \tilde G_{3/2} \,.
\end{align}
\end{subequations}
Hence, when supersymmetry is preserved, one is left with just one complex and two real independent form factors. 

One might like to require conformal invariance on top of
supersymmetry. The net effect on the form factors can be obtained by
observing that in such case $T=0$ as an operator and hence, by
supersymmetry, $X=0$. 
Let us notice that one could perform a shift \cite{Komargodski:2010rb} in the superfields $(\mathcal{J}_\mu,X)$ which leaves the definition \eqref{FZ} invariant. Here, choosing $X$ to be exactly equal to zero, we are fixing this ambiguity.
From now on we will always work within this assumption.
The vanishing of $X$ implies that 
\begin{equation}
\label{susyr}
F_2=F_{3/2}=F_1=F_0=0\, .
\end{equation}
As already observed, the vanishing of $X$ also implies that the non-conserved part of the two-point function of $j_\mu$ is projected out. Current conservation forces any correlator carrying a net charge under the  R-symmetry to vanish (notice that $R(X)=2$ and $R(S_\mu)=-1$). Hence also all complex form factors vanish in this case
\begin{equation}
\label{susyc}
H_2 = H_1 = G_{3/2} = \tilde G_{3/2}=0\, .
\end{equation}
Thus, in the superconformal case, only one (real) form factor survives. When conformal invariance is unbroken its functional dependence on $k^2$ is completely fixed up to an overall constant. This also shows that at a superconformal fixed point the central charge $c$ completely determines the two-point functions of the supercurrent and of the R-current, besides that of the energy-momentum tensor
\begin{equation}
C_2 = C_{3/2} = C_{1R}=\frac{c}{3\pi^2}\log\frac{\Lambda^2}{k^2}\, .\label{centralcharge}
\end{equation}

Eqs.~(\ref{susyr}) and (\ref{susyc}) give also an a posteriori justification for the presence of a mass scale in the parametrization of the traceful  part of real correlators and of the complex ones. Indeed, if the theory does not contain any scale, any correlator involving the mass scale $m$ should vanish.

The most generic situation is obtained in a supersymmetry breaking vacuum, where both $M$ and $m$ are necessarily different from zero and the form factors are not anymore related to one another, in general. Notice that since $T=0$ is an operator identity in a conformal theory, in order to break supersymmetry spontaneously and get a non-vanishing vacuum energy, conformal invariance must be explicitly broken. In other words, one can never have a situation in which $m=0$ and $M \not = 0$.

\subsection{Two-point functions for the $\mathcal{R}$ multiplet}

We now comment on the structure of two-point functions for the $\mathcal{ R}$ multiplet. Correlators not involving $C_{\mu\nu}$ have the same structure of those of the FZ multiplet (though the form factors will generically be different functions). One crucial difference, though, is that now $j_\mu$ is a conserved current and therefore $F_1=I_3=0$. 

As for correlators involving $C_{\mu\nu}$, the only non-vanishing ones are
\begin{subequations}
\begin{align}
&\langle C_{\mu\nu}(k)\,C_{\rho\sigma}(-k) \rangle =  3 \left(\eta_{\mu\rho} k_\nu k_\sigma - \eta_{\nu\rho} k_\mu k_\sigma + \eta_{\nu\sigma} k_\mu k_\rho -  \eta_{\mu\sigma} k_\nu k_\rho \right) m^2 E_0(k^2) \label{<CC>}  \\ 
&\langle C_{\mu\nu}(k)\,j_\rho(-k) \rangle = \frac i2 \left(\eta_{\mu\rho} k_\nu - \eta_{\nu\rho} k_\mu \right) m^2 E_1(k^2) \,,\label{<Cj>} 
\end{align}
\end{subequations}
where $E_0$ and $E_1$ are real form factors, and numerical coefficients have been chosen for later convenience.

One can easily work out the supersymmetry transformations of the fields belonging to the ${\mathcal R}$ multiplet, and find that in a supersymmetric vacuum the following relations between form factors should hold
\begin{equation}\label{SUSYrelR}
C_2=C_{3/2}=C_{1R}\quad,\quad F_2=F_{3/2} = E_1 = E_0\,.
\end{equation}
So, in this case, one is left with two independent real form factors. Notice the difference with respect to the FZ multiplet, for which the R-current is not conserved and, in turn, there can be a non-vanishing complex form factor in a supersymmetric vacuum, see eq.~(\ref{SUSYcomp2}). 
For  ease of notation,  in (\ref{SUSYrelR}) we have used the same
letters adopted for the FZ multiplet for correlators  involving
$T_{\mu\nu}$, $S_{\mu\alpha}$ or $j_\mu$, but the explicit form of the $F_s$ and $C_s$  is a priori different.

For a superconformal theory, the ${\mathcal R}$ and FZ multiplets can be
chosen to coincide by selecting the superconformal R-current as the
bottom component of ${\mathcal R_\mu}$. In this case, one finds that
$F_2=F_{3/2}=E_1=E_0= 0$, while $C_{1R}=C_{3/2}=C_2 \not = 0$, as for
the FZ multiplet, and one is consistently left with only one real form
factor. However, in the context of R-symmetric RG flows, there is another natural choice for the lowest component of $\mathcal R_\mu$ at the UV fixed point, that is to select the R-symmetry preserved along the flow (let us assume for simplicity that it is unique). In this case, at the UV and IR fixed points one gets 
\be
F_2 = F_{3/2} = E_1 = E_0 = \frac13 \frac{k^2}{m^2} \frac{1}{(2\pi)^2} \tau_U^{UV,\,IR} \log\frac{\Lambda^2}{k^2}.\label{tauBuican}
\ee
The quantities $\tau_U^{UV}$ and $\tau_U^{IR}$ have been studied in \cite{Buican:2011ty}, where they were conjectured to satisfy the inequality $\tau_U^{UV}>\tau_U^{IR}$.

\section{Supercurrent correlators at weak coupling}

In this section, as a warm-up, we will apply the formalism we have introduced to the simplest class of weakly coupled models one can think of, that is  WZ-like models with a single chiral superfield. By considering a free massless chiral superfield, a massive one and finally a model with a linear superpotential perturbation (i.e.~the Polonyi model), we will compute the explicit form of the form factors discussed in the previous section in toy-examples of superconformal theories (both in vacua preserving and not preserving conformal invariance), supersymmetric but not conformal theories and, finally, theories breaking supersymmetry spontaneously. 

\subsection{Conformal case: massless chiral multiplet}
\label{sectionmassless}

Let us consider a theory of one chiral massless superfield $\Phi$ with canonical K\"ahler potential and no superpotential, that is a free theory. The FZ multiplet is given by (see, e.g.~\cite{Komargodski:2010rb})
\begin{subequations}
\begin{align}
 \mathcal J_\mu & = -\frac 1 6 \left(\bar D  \Phi^*\right) \bar\sigma_\mu \left(D \Phi\right) + \frac 2 3 i\left(\Phi^*\partial_\mu\Phi-\Phi\partial_\mu\Phi^*  \right) \label{JWZfree}
\\ X & = -\frac1 3 \Phi\bar D^2 \Phi^*\,.
\end{align}
\end{subequations}
The equations of motion for $\Phi$ are simply
\begin{equation}
 D^2 \Phi = \bar D^2 \Phi^* = 0\,,
\end{equation}
thus we see that on-shell $X=0$, as appropriate for a superconformal field theory. 

From the component expression of the FZ superfield \eqref{JWZfree} we get
\begin{subequations}
\begin{align}
\timn = & \frac1 3\left( 4\,\partial_{(\mu}\phi\partial_{\nu)}\phi^*- \nmn\partial^\rho\phi\partial_\rho\phi^* -  \phi^*\partial_\mu\partial_\nu\phi -  \phi\partial_\mu\partial_\nu\phi^*\right)\label{TWZfree}\nonumber \\ 
           & 
           +\frac i 2 \left(\, \psi\sigma_{(\mu}\partial_{\nu)}\bar\psi -\partial_{(\mu}\psi\sigma_{\nu)}\bar\psi\, \right)
\\
   \sma = & \frac{2\sqrt2}{3}i\left( \phi^*\partial_\mu\psi_\alpha - \psi_\alpha\partial_\mu\phi^* + \frac1 2 (\sigma_\rho\bar\sigma_\mu)_\alpha^{\phantom\alpha\beta}\psi_\beta\partial^\rho\phi^* \right)
   \label{SWZfree}
\\  j_\mu = &\frac1 3 \psi\sigma_\mu\bar\psi + \frac2 3 i\left( \phi^*\partial_\mu\phi-\phi\,\partial_\mu\phi^* \right) \,,\label{jWZfree}
\end{align}
\end{subequations}
where we have neglected terms which vanish on-shell, and we have used the usual parametri\-zation for a chiral superfield, i.e.  
$\Phi(x,\theta,\thetab) = \phi + \sqrt2 \theta\psi + \theta^2 F +\dots$ where the ellipses stand for the supersymmetric completion. 

\begin{figure}
\centering
\includegraphics[scale=0.56]{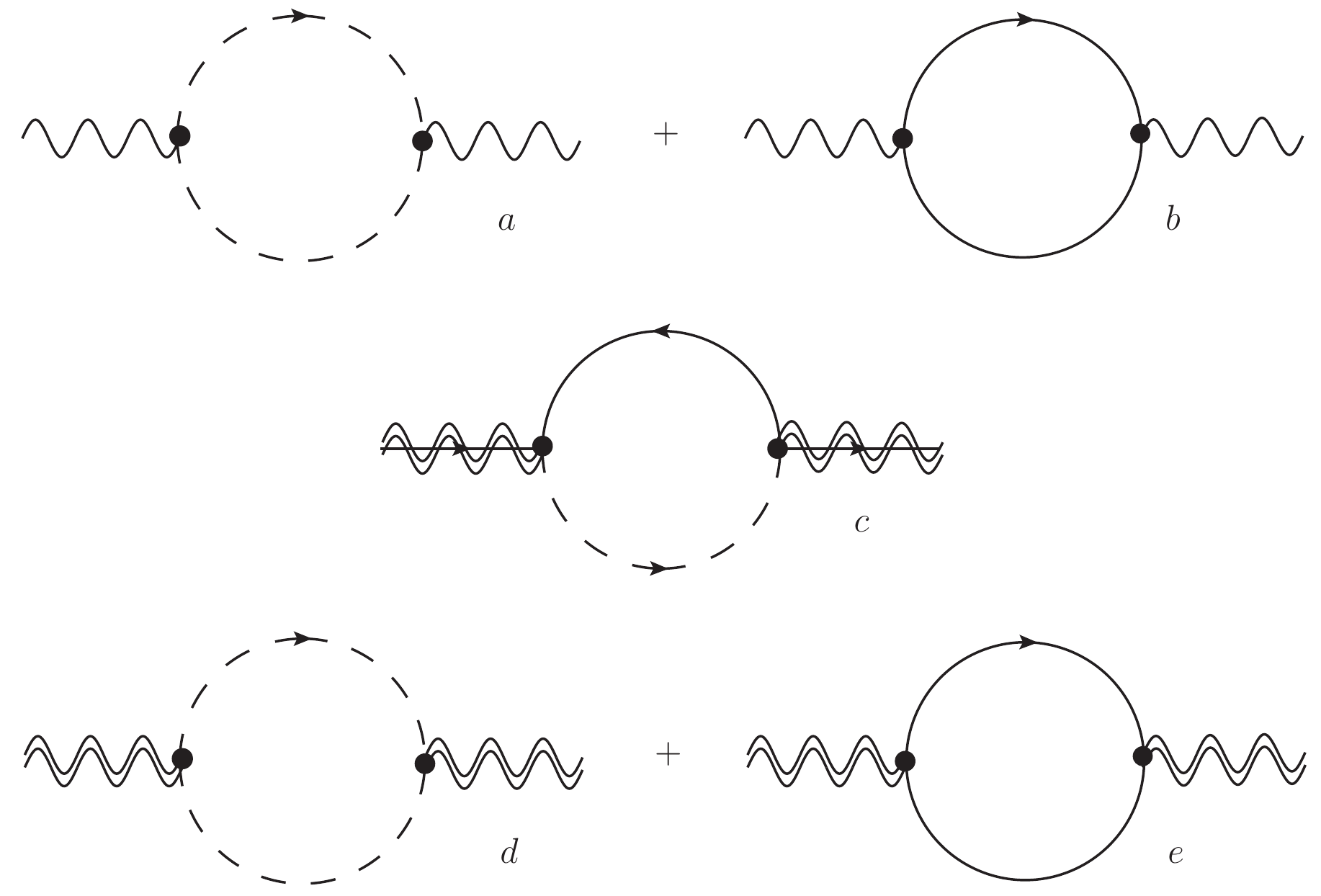}
\caption{The one-loop contributions to $\langle j_\mu j_\nu \rangle$:  $(a+b)$,  $\langle \sma \bar S_{\nu\dot\beta} \rangle$: $c$ and $\langle \timn T_{\rho\sigma} \rangle$: $(d+e)$. The external lines are meant to represent the source (supergravity) fields for each operator.}\label{fig:2}
\end{figure}

One can easily check that $T$, $S$ and $\partial_\mu j^\mu$ are all zero on-shell. As expected for a superconformal theory, some of the real form factors and all the complex ones  vanish in this case
\begin{equation}
G_{3/2} = \tilde G_{3/2}= H_2 = H_1 = I_3=0\,,\,   F_2 = F_{3/2} = F_1=F_0 = 0\,.\label{confcomp}
\end{equation}
So we are left with the computation of the traceless part of the correlators (\ref{<TT>}), (\ref{<SSbar2>}) and (\ref{<jj>}), namely of $C_2$, $C_{3/2}$ and $C_{1R}$. This can be done by evaluating the one-loop diagrams in Fig.~\ref{fig:2}. In what follows we discuss separately the cases where the vacuum preserves or does not preserve the superconformal symmetry.

\subsubsection*{\it Unbroken conformal invariance}

Evaluating the one-loop diagrams of Fig. \ref{fig:2}, one gets the following result
\begin{equation}
C_2(k^2) =C_{3/2}(k^2) = C_{1R}(k^2) =  \frac{1}{(4\pi)^2} \frac29 \left( \log\frac{\Lambda^2}{k^2} + \frac73 \right)\,.
\end{equation}
As expected, the three form factors are equal and have the correct
logarithmic behavior for a conformal theory. In particular, comparing
our result with (\ref{centralcharge}), we get the value of the central
charge $c=1/24$ which is in agreement with the expected result for a
free theory with one chiral multiplet \cite{Anselmi:1997am}.  

In evaluating the one-loop diagrams we have used dimensional
regularization in the minimal subtraction scheme. Terms which are
polynomial in the external momentum can of course be scheme-dependent,
and in that case they do not correspond to physical observables. For
instance, in our scheme there is a finite contribution such that
$F_{2}=F_{3/2}=\frac{1}{4\pi^2}\frac{2}{9}\frac{k^2}{m^2}\neq F_1
= 0$. This is not surpising, because  dimensional regularization does
neither preserve conformal symmetry nor supersymmetry, as it is
evident considering that in $d=4-2\epsilon$ dimensions $T \neq0$ and
$S\neq0$ while it remains true that
$\partial_{\mu}j^{\mu}=0$.\footnote{A contact-term contribution to
  $F_{2}$ in a CFT is related to the $a'$ coefficient in the trace of
  the energy-momentum tensor on a curved background, first discussed
  in \cite{Osborn:1993cr} (where the coefficient is dubbed $h$)  and
  then extended to the supersymmetric case in
  \cite{Bonora:1984pn,Buchbinder:1986im,Osborn:1998qu}. Since it can
  be shifted by adding a local counterterm, this coefficient is not a
  real anomaly, neither its value can be considered as a datum of the
  CFT. See however \cite{Cappelli:1990yc,Anselmi:1999xk,Anselmi:2002mk} for a
  tentative interpretation of the difference $a'_{UV}-a'_{IR}$ in the
  presence of an RG-flow.} For instance, using differential
regularization \cite{Freedman:1991tk}, both the residual constant terms
in the $C_s$ form factors and  the $F_s$ form factors can be shown to
vanish. Anyhow, since they do not play any important role in our
discussion, we will not be concerned with contact terms from now on.

\subsubsection*{\it Spontaneous breaking of conformal invariance}

We now want to consider situations in which conformal symmetry (and hence, by supersymmetry, also the superconformal R-symmetry) is broken spontaneously. Sticking to our simple toy-model, this can be achieved by choosing a non-zero VEV for the lowest component of the chiral superfield $\Phi$, that is $\langle \phi\rangle=v$.\footnote{Note that two vacua with a different value of $v$ are not really physically different in this simple model, because of an additional symmetry shifting the superfield $\Phi$ by a constant. This additional spurious symmetry can be removed considering a model closer in spirit to, e.g., the Coulomb branch of ${\mathcal N}=4$ SYM. However, the poles that one would recover will be analogous to those in our single field model, to which we then focus for simplicity. 
}

Since conformal invariance is broken spontaneously, the operator identity $X=0$ still holds, on-shell. Hence all correlators that were vanishing before are still vanishing, see eqs.~(\ref{confcomp}). On the other hand, we should  now find   poles in the traceless part of the correlators, corresponding to the Goldstone modes associated to the broken symmetries. Since supersymmetry is not broken, Goldstone modes should appear in supermultiplets. Therefore, we expect a dilaton, an axino/dilatino, and an R-axion in this case, which correspond to poles in $C_2, C_{3/2}$ and $C_{1R}$, respectively. 

A simplification, in what follows, is that in order to find the poles, it will be sufficient to determine the piece of the current operators which is linear in the fields, and then compute the correlators at tree level (the loop parts will be closely related to the ones discussed in the conformal case). We thus start by finding the linear pieces in the operators \eqref{TWZfree}, \eqref{SWZfree} and \eqref{jWZfree} which read
\begin{subequations}
\begin{align}
T_{\mu\nu}^\mathrm{lin} = &  -\frac{1}{3} v \partial_\mu \partial_\nu(\phi+\phi^*)  \\ 
S_\mu^\mathrm{lin}= & \frac{2\sqrt{2}}{3}i v \partial_\mu \psi  \\
j_\mu^\mathrm{lin} = & \frac{2}{3}i v \partial_\mu (\phi-\phi^*)\, ,
\end{align}
\end{subequations}
where we have taken $v$ to be real, for simplicity. 

Let us start with the tree-level correlator of $j_\mu^\mathrm{lin}$, which can be easily evaluated to be
\begin{equation}
\langle j_\mu^\mathrm{lin}(k)j_\nu^\mathrm{lin}(-k)\rangle = \frac{8}{9} v^2 k_\mu k_\nu \frac{1}{k^2}\, .
\end{equation}
The trasversality can be restored by introducing the seagull-like term familiar in scalar QED, and adding the corresponding contact term $-\frac{8}{9} v^2 \eta_{\mu\nu}$ to the tree-level correlator. The end result we get for the form factor reads
\begin{equation}
\label{C1rb}
C_{1R} (k^2)= \frac{8}{9} \frac{v^2}{k^2}\ ,
\end{equation}
correctly displaying the pole associated  to the R-axion.

The supercurrent and energy-momentum tensor correlators can be equivalently evaluated and read
\begin{subequations}
\begin{align}
\langle S_\mu^\mathrm{lin}(k)\bar S_\nu^\mathrm{lin}(-k)\rangle & = - \frac{8}{9} \frac{v^2}{k^2}\,Y_{\mu\nu}  =  \frac 89 \frac{v^2}{k^2} k_\mu k_\nu k_\rho \sigma^\rho + \mathrm{contact\;terms} \, , \\
\langle T_{\mu\nu}^\mathrm{lin}(k)T_{\rho\sigma}^\mathrm{lin}(-k)\rangle & =  - \frac 19 \frac{v^2}{k^2}X_{\mu\nu\rho\sigma}  =  - \frac 19 \frac{v^2}{k^2}k_\mu k_\nu k_\rho k_\sigma + \mathrm{contact\;terms}\,,
\end{align}
\end{subequations}
where again, contact terms have been added to get a transverse result, and $X_{\mu\nu\rho\sigma}$ and $Y_{\mu\nu}$ are defined in eqs.~(\ref{X}) and (\ref{Y}). These correlators lead to form factors which are identical to (\ref{C1rb}), and the corresponding poles are associated to the axino/dilatino and the dilaton, respectively.

Let us finally remark that the contact terms we have added to restore transversality for the supercurrent and the energy-momentum tensor can be put in correspondence to quadratic local terms in the sources which are present in the supergravity lagrangian \cite{Wess:1992cp}, in analogy with the seagull for the vector field.  

\subsection{Supersymmetry without conformal invariance: massive chiral multiplet}
\label{sectionmassive}

We will now consider a case in which a mass term is added, $W = \frac 12 m \Phi^2$. This superpotential term, while breaking the superconformal R-symmetry together with conformal symmetry, still allows for a different R-symmetry, under which $R(\Phi)=1$. The superfield definition for $\mathcal{J}_\mu$ is unchanged with respect to the conformal case given in (\ref{JWZfree}), but the final expression for the component operators contains new terms, because it is obtained by using the modified equations of motion. The result is
\begin{subequations}
\begin{align}
 \timn = &   \frac1 3 \left(4\partial_{(\mu}\phi\partial_{\nu)}\phi^* -  \phi^*\partial_\mu\partial_\nu\phi  -  \phi\partial_\mu\partial_\nu\phi^*-\nmn\partial^\rho\phi\partial_\rho\phi^* -\nmn |m\phi|^2\right)\label{massivestress}
 \\\nonumber & +\frac i 2 \left(\psi\sigma_{(\mu}\partial_{\nu)}\bar\psi -\partial_{(\mu}\psi\sigma_{\nu)}\bar\psi\right)
 \, \\
    \sma = & \frac{2\sqrt2}{3}i\left( \phi^*\partial_\mu\psi_\alpha - \psi_\alpha\partial_\mu\phi^* + \frac1 2 (\sigma_\rho\bar\sigma_\mu)_\alpha^{\phantom\alpha\beta}\psi_\beta\partial^\rho\phi^* \right) + \frac{\sqrt2}{3} \sigma^\mu \bar\psi m^*\phi^*\, \label{supermassive}.
\end{align}
\end{subequations}
The presence of a superpotential introduces a new term in the expression for $X$
\begin{align}
X & = 4 W -\frac1 3 \Phi\bar D^2 \Phi^*\, \label{XWZ}\,.
\end{align}
which, by using the equations of motion becomes
\begin{equation}
 X =  \frac23\, m\, \Phi^2 =  \frac23 m \phi^2 + \frac23 \theta\left( 2\sqrt2 m \phi\psi \right) -\frac43 \theta^2\left( |m\phi|^2 + \frac12 m \psi^2 \right)\,.
\end{equation}
Comparing the above expression with \eqref{X-comp} we can easily read the expressions of the trace operators
\begin{subequations}\label{traces}
 \begin{align}
 & T = - 2 |m\phi|^2 - \left( \frac12 m \psi^2  + c.c.\right)
\\& S = 2\sqrt2 m \phi\psi
\\& \partial_\mu j^\mu = i\frac13\left( m \psi^2 - c.c \right)
\\ & x = \frac23 m \phi^2\,.
 \end{align}
\end{subequations}

The main difference between this example and the previous ones is that, since conformal symmetry is explicitly broken, we expect one more real form factor to be generated (the complex one, which generically can be non-zero in the non-conformal case, is forbidden by the unbroken R-symmetry). This is the form factor corresponding to the traceful part of the correlators, namely $F_2$, $F_{3/2}$ and $F_1$ in our parametrization, which are predicted to be equal when supersymmetry is unbroken. 

\begin{figure}
\centering
\includegraphics[scale=0.56]{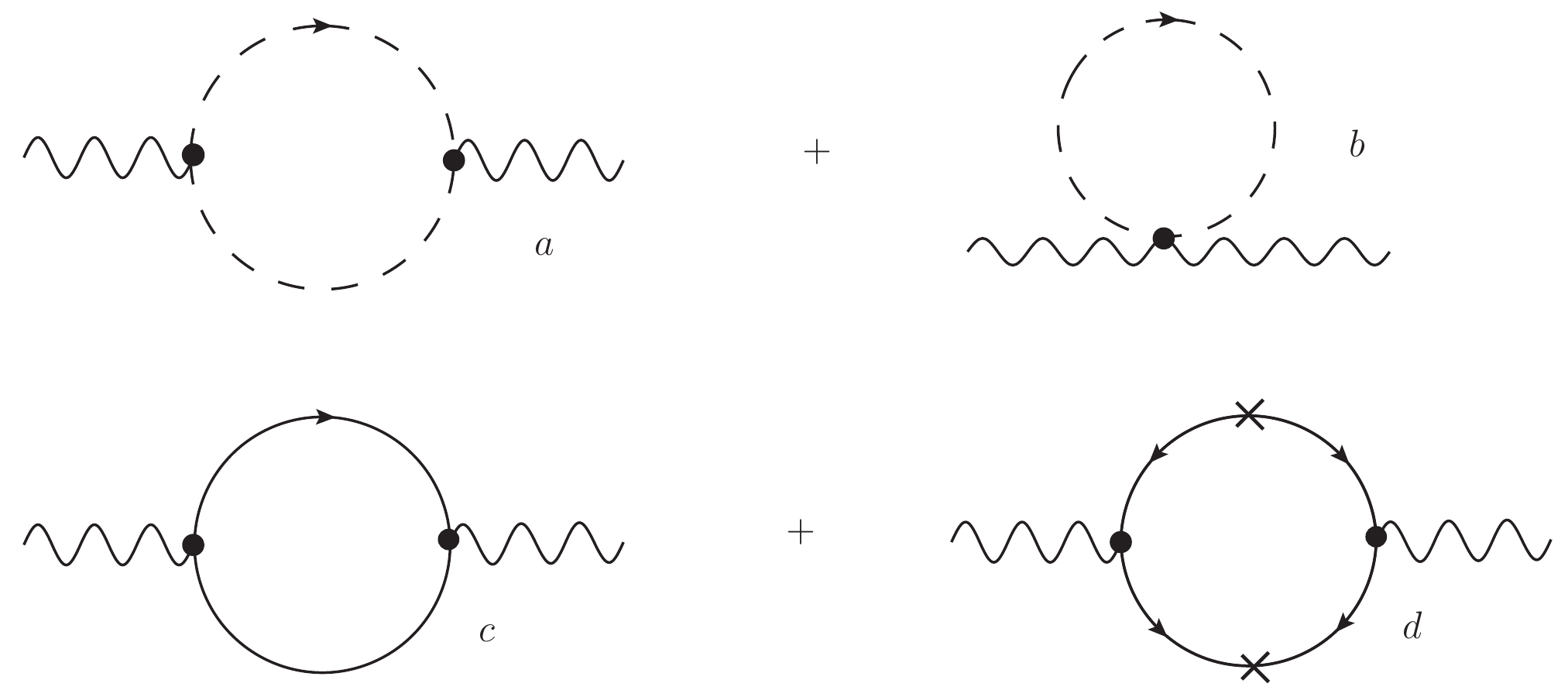}
\caption{The one-loop contributions to $\langle j_\mu j_\nu \rangle$ in the massive case. Diagram $b$ is a contact term needed for (bosonic) current conservation. For the ${\mathcal R}$ multiplet, only bosonic loops contribute, since $R(\Phi)=1$ and hence $R(\psi)=0$.}\label{fig:3}
\end{figure}

Evaluating the one-loop diagrams of Fig. \ref{fig:3}, one gets the following result for the corresponding form factors
\begin{subequations}
\begin{align}
C_{1\tilde{R}} & = \frac{1}{(4\pi)^2}\frac 29 \left(\log\frac{\Lambda^2}{m^2} + \frac 73 + \frac {4m^2}{k^2} - \frac{k^2+2\,m^2}{k^2} \mathcal{L}\left(\frac{k^2}{m^2}\right)\right) \label{C1FZ}\\
F_1 & = \frac{1}{(4\pi)^2}\frac 43 \left(\log\frac{\Lambda^2}{m^2} + 2 -  \mathcal{L}\left(\frac{k^2}{m^2}\right)\right)\,, \label{F1m}
\end{align}
\end{subequations}
where
\be
\mathcal{L}(x) = 2\, \sqrt{\frac{4+x}{x}}\,\mathrm{arctanh}\left(\sqrt{\frac{x}{4+x}}\right)\, 
\ee
and $\tilde R$ stands for the non-conserved superconformal
R-current. For $k^2\ll m^2$, both $C_{1\tilde{R}}$ and $F_1$ go to a
finite constant, which is what we expect from a theory with a mass gap.

Note that expanding the above expression for
$C_{1\tilde{R}}$ at fixed $k^2$ for $m$ going to zero we
approach the free conformal fixed point and we recover the central
charge $c=\frac{1}{24}$ as the coefficient of the leading term in the
series. Conversely, for  $m$ approaching the cut-off $\Lambda$ $C_{1\tilde{R}}$ goes to zero, in agreement with the fact that the end of the flow is the empty theory. $C_{1\tilde{R}}$ is indeed a good candidate for being a central function which interpolates between the central charges of two superconformal field theories connected by an RG flow \cite{Anselmi:1997am}.

Let us finally notice that when continued in (mostly plus) lorentzian signature, there is a branch cut for $- k^2 > 4m^2$ associated to the multi-particle channel and also that only the fermion loops contribute to $F_1$, since the bosonic part of the R-current is still conserved.

The supersymmetric relation between non-conformal form factors can be verified by explicitly calculating the two-point functions of the energy-momentum tensor and of the supercurrent operators in the presence of the mass perturbation. A simpler way is to rely on the relation between the traces and the mass operator, as explained in Appendix  \ref{OWI}. In this case this amounts to substituting the relations ($\ref{traces}$) in the correlators ($\ref{non-conformal}$), and one can easily see that indeed $F_2 = F_{ 3/2} = F_1$.

Since there is a conserved R-symmetry in this case, it is also
possible to define an $\mathcal{R}$ multiplet, and it is instructive to see how the two-point correlators of this multiplet behave in this simple case, comparing the results with the ones obtained for the FZ multiplet. 

Starting from the general superfield definition of the $\mathcal{R}$ multiplet for a theory in which also the FZ multiplet is well defined, it is easy to see that 
\begin{equation}
\mathcal{R}_{\mu}=i\left(\Phi^{\ast}\partial_{\mu}\Phi-\Phi\partial_{\mu}\Phi^{\ast}\right)\ ,
\qquad 
\chi_{\alpha}=-\frac{1}{2}\bar{D}^2D_{\alpha}\left(\Phi\Phi^{\ast}\right)\ ,
\end{equation}
where the conserved current sitting in the bottom component of the $\mathcal{R}$ multiplet is indeed associated with the R-symmetry under which $R[\Phi]=1$. 

From the component expression of the $\mathcal{R}$ multiplet ($\ref{R-comp}$) we derive that 
\begin{subequations}
\begin{align}
T_{\mu\nu}=&\ \frac{1}{2}\left(\partial_{\mu}\phi\partial_{\nu}\phi^{\ast}+\partial_{\nu}\phi\partial_{\mu}\phi^{\ast}-\phi\partial_{\mu}\partial_{\nu}\phi^{\ast}-\phi^{\ast}\partial_{\mu}\partial_{\nu}\phi\right)
 \\\nonumber &+\frac i 2 \left(\, \psi\sigma_{(\mu}\partial_{\nu)}\bar\psi -\partial_{(\mu}\psi\sigma_{\nu)}\bar\psi\, \right)\\
S_{\mu\alpha}=&\ i\sqrt{2}(\phi^{\ast}\partial_{\mu}\psi_{\alpha}-\psi_{\alpha}\partial_{\mu}\phi^{\ast}) \\
j_{\mu}=&\ i(\phi^{\ast}\partial_{\mu}\phi-\phi\partial_{\mu}\phi^{\ast})\, ,
\end{align}
\end{subequations}
where  $j_{\mu}$ only depends on the scalar (consistently with the fact that $R(\psi)=0$), while $T_{\mu\nu}$ and $S_{\mu\alpha}$ are obtained from the FZ ones in (\ref{massivestress})--(\ref{supermassive}) via an improvement transformation of the type (\ref{shift1}) with $U =-\frac 13 \Phi\Phi^* $.  
Since the current is conserved, we expect a transverse correlator determined by only one form factor which is
\be
\label{C1Rm}
C_{1R} = \frac{1}{(4\pi)^2} \frac 13 \left( \log\frac{\Lambda^2}{m^2} + \frac 8 3 + \frac {8 \, m^2}{k^2} - \frac{k^2+4\,m^2}{k^2} \mathcal{L}\left(\frac{k^2}{m^2}\right)\right)\, . 
\ee
Even if similar in form, the result (\ref{C1Rm}) differs from the analogous one for the FZ multiplet (\ref{C1FZ}). In particular the leading term in the UV expansion gives the central charge associated with a $U(1)_{R}$ which is not the superconformal R-symmetry and therefore cannot be identified with the central charge $c$. 

Since the conserved R-current depends only on the scalar, the mixed correlator with $C_{\mu\nu}$ ends up to be proportional to $C_{1R}$. In particular we get
\begin{equation}
E_{1}=\frac{k^2}{m^2}C_{1R}\ .\label{E1massive}
\end{equation}
Taking $k^2$ fixed and inserting the form factor (\ref{E1massive}) in
the correlator (\ref{<Cj>}), we see that the latter does not go to
zero in the limit in which $m^2$ goes to zero. The coefficient of the
logarithmic divergence gives the $\tau_U$ quantity defined in
(\ref{tauBuican}), as expected.

\subsection{Spontaneous supersymmetry breaking}
\label{ssb}

Finally, we want to consider a case where supersymmetry is (spontaneously) broken. The simplest such model is the Polonyi model, which amounts to add a linear potential to the theory of a free massless chiral superfield of section \ref{sectionmassless} 
\begin{equation}
W = f \Phi\,.
\end{equation}
In the vacuum of the Polonyi model we have that $\langle F\rangle =-f^*$, while the one-point function of $T_{\mu\nu}$ is simply given by
$\langle T_{\mu\nu}\rangle =- |f|^2 \eta_{\mu\nu}$, 
signalling the spontaneous breaking of supersymmetry. This implies the existence of a Goldstino in the spectrum, which, as discussed in section \ref{formfactors}, should show up as a pole in the two-point function (\ref{<SSbar2>}). Finally, form factors are not expected anymore to respect equalities like those in  (\ref{SUSYcomp}) and (\ref{SUSYcomp2}), since supersymmetry is broken.
 
In our present model we have that
\be
\label{Xsb}
X= \frac 83 f\Phi  = \frac 83 f \left( \phi + \sqrt2 \theta\psi - \theta^2 f^* +  \dots\right) \,,
\ee
which implies, by comparison with eq.~(\ref{X-comp}), that at linear level the supercurrent $S_{\mu\alpha}$ reads 
\begin{align}
S_{\mu\alpha}^\mathrm{lin}= &\sqrt{2}f^* (\sigma_\mu \bar \psi)_\alpha\,.
\end{align}
From this expression, the real fermionic correlator  can be easily computed to be
\begin{align}
\langle S_{\mu\alpha}^\mathrm{lin}(k) \bar S_{\nu\dot\beta}^\mathrm{lin}(-k)\rangle&=2|f|^2 \sigma_\mu \bar\sigma^\rho \sigma_\nu \frac{k_\rho}{k^2} \, ,
\end{align}
which, by setting $|f|= M^2$, agrees with the expected expression (\ref{<SSbar2>}) and the existence of a Goldstino mode. 

At linear level there is no contribution to the form factors $C_{3/2}$ and $F_{3/2}$. In fact, due to the simplicity of the superpotential perturbation, the supersymmetry breaking deformation does not affect the one-loop computation of the form factors, and one gets the same result as for the superconformal case. In particular, $F_{3/2}=0$. The same holds for all other real correlators, hence $F_1=F_2=0$. This could seem at odds with the fact that the theory breaks supersymmetry, but is in fact the price to pay for having chosen such a simple model (which is a free theory). The violation of the supersymmetric relations (\ref{SUSYcomp}), though, can be seen by computing $F_0$, which does get a contribution at linear level and is different from zero. From eq. (\ref{Xsb}) one easily gets
\begin{equation}
\langle x (k)x^*(-k)\rangle = \frac{64}{9}|f|^2 \frac{1}{k^2} \not = 0\, .
\end{equation}
The appearance of a pole is due to the presence of a massless sGoldstino (the pseudomoduli space is not lifted quantum mechanically in the Polonyi model). Such massless mode can be lifted if one considers, e.g., a model where the complex boson is given a mass or is  removed from the spectrum, as by imposing the constraint $\Phi^2=0$.

\section{Supercurrent correlators at strong coupling}

In this section we consider theories at strong coupling, and compute correlators of the supercurrent multiplet 
using holography. According to the field/operator correspondence  \cite{Gubser:1998bc,Witten:1998qj}, the bulk fields dual to the stress-energy tensor, the supercurrent and the R-current are the graviton, the gravitino and the graviphoton, respectively. Hence, in order to get  two-point functions of the supercurrent  multiplet
we will consider linearized fluctuations of the graviton multiplet. 

As we did in the previous section, we will stick, in what follows, to  
the simplest possible set-up, namely a field theory whose gravity dual is
Anti-de Sitter space-time possibly cut-off by a hard wall in the
bulk \cite{ArkaniHamed:2000ds,Rattazzi:2000hs,Polchinski:2001tt,BoschiFilho:2002vd,Erlich:2005qh}. 
This is a bottom-up model, which
is however flexible enough to let one reproduce most of the physics
we discussed in the weakly coupled case. The background is described by an AdS metric which can be written as
\be
ds^2= \frac{1}{z^2}\big( dz^2+\eta_{\mu\nu}dx^\mu dx^\nu\big) \, , 
\label{ads}
\ee
understood to be extending from the boundary at $z=0$ to a cut-off at
$z=1/\mu$, which geometrically is indeed a hard wall. The boundary $z=0$
corresponds to the deep UV of the quantum field theory, while the cut-off $z=1/\mu$ represents the
smallest scale in the IR, here given by $\mu$.  

Locally, for all values
of $z$ larger than the IR cut-off, the whole (conformal) isometry
group of AdS is unbroken. Thus a hard wall is a (very simplified) model for a theory which flows from a UV conformal
fixed point to a gapped phase in the IR, with spontaneously broken conformal symmetry \cite{ArkaniHamed:2000ds,Rattazzi:2000hs}. On 
the contrary, one recovers a fully conformal field
theory when $\mu \to 0$ and AdS space-time is no longer cut-off. 
Indeed, by considering the fluctuations of the graviton, the gravitino and the graviphoton, and applying the 
standard AdS/CFT machinery, we will see that one gets the correlators of a SCFT in unbroken and broken phases, 
for $\mu=0$ and $\mu \not = 0$, respectively. In particular, in the latter case, we will show that $1/k^2$ poles arise in the form factors, 
corresponding to massless dilaton, dilatino and R-axion.

In theories where conformal symmetry is explicitly broken, $X \not = 0$. In this case, the graviton multiplet does not have enough degrees of freedom to describe, holographically, the FZ multiplet (in particular, one cannot generate non-trivial $F_s$ form factors), and at least one hypermultiplet, dual to $X$, must be added.\footnote{Completely analogous statements can be made for the ${\mathcal R}$ multiplet, where the extra fields sit in a vector multiplet dual to the real superfield $U$ (or in a tensor multiplet dual to $\chi_\alpha$, in theories where the FZ multiplet is not defined).} 

This agrees with the fact that specific non-trivial profiles of scalar fields are needed in order to describe, holographically, non-conformal theories, the scalar being dual to the operator perturbing the fixed point. One should then consider the backreacted solution for the coupled system given by the scalar and the metric (and possibly their supersymmetric partners). This implies that the hard wall is a too simple background to describe field theories in which conformal invariance is explicitly broken and, eventually, theories with spontaneously broken supersymmetry. The analysis of richer backgrounds, with fully backreacted scalar profiles, is left for future work. Here we will take an effective approach, which consists in working at the lowest order in the relevant perturbation of the fixed point. The basic idea is that we start with the conformal theory in the non-conformal vacuum parametrized by the scale $\mu$ of the IR wall, and then treat a perturbation with relevant coupling $m$, in an expansion in 
$ m/\mu$. By means of the Ward identities of (broken) conformal invariance,  eqs.~(\ref{XO}) and (\ref{traces-perturbation1})-(\ref{traces-perturbation4}), this will allow us to recover the non-conformal form factors $F_s$ at lowest order in this expansion, simply by considering fluctuations of the hypermultiplet on the un-backreacted hard wall background. This same short-cut approach will enable us to describe, holographically, supersymmetry breaking models and get, in particular, the expected Goldstino pole in $\langle \bar S_{\dot \alpha} S_\alpha\rangle $.

In what follows, correlators are computed  through the
procedure of holographic renormalization
\cite{deHaro:2000xn,Bianchi:2001de,
  Bianchi:2001kw,Skenderis:2002wp}. These are by now standard
techniques, hence we will not go into any technicality in the
remainder of this section, and just discuss the results we
obtain. However, several useful technical details of the procedure,
specialized to the hard wall background, are presented in appendix
\ref{appholo}, to which the interested reader can refer to. We will
always set our  computations in the framework of ${\mathcal
  N}=2$ gauged supergravity, and exploit the holographic dictionary to
compute correlators at the complete supermultiplet level, as
initiated in  \cite{Argurio:2012cd,Argurio:2012bi}. This is a necessary ingredient in order to deal with strongly coupled supersymmetric QFT systematically, and have control on their (supersymmetry breaking) dynamics.

\subsection{Unbroken conformal symmetry}

We start by the most symmetric case, which amounts to consider 
fluctuations of the graviton supermultiplet on a pure AdS$_5$
background. This multiplet contains the graviton $h_{MN}$, the gravitino $\psi_M$ and
the graviphoton $A_M$, and the corresponding action is
\be
S=\frac{N^2}{4\pi^2}\int \text{d}^5x\sqrt{G}\Big(-\frac12 R-6
+\bar{\psi}_M(\Gamma^{MNP}{D}_N-\frac32\Gamma^{MP})\psi_P+
\frac14 G^{MP}G^{NQ}F_{MN}F_{PQ}\Big)\, , \label{bulkaction}
\ee
where we have not written boundary terms. The five dimensional Newton
constant is fixed in terms of the $AdS_5\times S^5$ ten-dimensional solution, taking $L_\mathrm{AdS}=\alpha'=1$,
and we use indices such that $x^M=(z,x^\mu)$. 
Unbroken conformal symmetry
implies, by supersymmetry, also unbroken superconformal R-symmetry,
so that, consistently, the graviphoton is massless. Also, supersymmetry
in AdS implies the gravitino has mass $|m|=\frac32$,  in units of
the AdS radius. 

In order to compute two-point correlators, we need to consider only quadratic
fluctuations of the bulk fields. In this simple set-up, we can restrain to fluctuations that are
completely gauge-fixed, $h_{Mz}=A_z=\psi_z=0$. We can furthermore consider
transverse and traceless $h_{\mu\nu}$, transverse and
$\Gamma$-traceless $\psi_\mu$, and transverse $A_\mu$. 

The essence of extracting correlators holographically is the following.
In a near-boundary expansions, fluctuations have two independent modes,
one leading and one sub-leading, that determine the whole
solution. Regularity conditions in the deep interior of AdS or
boundary conditions at the hard-wall then fix the dependence of the
subleading mode in terms of the leading one. The two-point correlator
is precisely given by this dependence, up to some local contact terms that can be set to zero in a suitable subtraction scheme (see appendix  \ref{appholo} for details).

In pure AdS, the bulk condition is that the fluctuation does not
explode in the deep interior. This fixes the solution uniquely, and going through all the
procedure of holographic renormalization one gets the correlators (\ref{<TT>})-(\ref{<jj>}), expressed in terms of the following
form factors
\begin{equation}
C_{2}(k^2)=C_{3/2}(k^2)=C_{1R}(k^2)=C^{AdS}(k^2)=\frac{N^2}{12\pi^2}\log\frac{\Lambda^2}{k^2}\, ,
\end{equation}
where $\Lambda$ is a UV regulator, and there can be additional
constant pieces according to the subtraction scheme (see appendix \ref{appholo}).  All other form factors vanish. These results are the expected ones for a superconformal field theory.
In particular, the value for $C_2$ is the well-known result
\cite{Gubser:1998bc} of the holographic
derivation of the central charge of ${\mathcal N}=4$ SYM, for which
$c=a=\frac{N^2}{4}$ in the large $N$ limit. What we have explicitly shown here
is that the same central charge is recovered from the
R-current correlator and from the supercurrent correlator, consistently with supersymmetry and eq. (\ref{SUSYcomp}).

\subsection{Spontaneously broken conformal symmetry}

In order to reproduce a situation where the field theory has a
vacuum where conformal symmetry is spontaneously broken, we consider
AdS space-time cut-off at $z=1/\mu$ where the scale $\mu$
is identified with the scale of the VEV that breaks the conformal symmetry. The
hard wall is modeling a theory where such spontaneous breaking leads to
a discrete spectrum, typical of a confining theory. 

Differently from pure AdS, the
geometry now ends abruptly at the wall  $z=1/\mu$, and we have to impose
there generic homogeneous boundary conditions for the field fluctuations 
\begin{subequations}
\begin{align}
&(h_{\mu\nu}(z,k)+\rho_{2}z\partial_{z}h_{\mu\nu}(z,k))\vert_{z=1/\mu}=0 \label{bch}\\
&(\psi^\mu(z,k)+\rho_{3/2}z\partial_{z}\psi^\mu(z,k))\vert_{z=1/\mu}=0 \label{bcpsi} \\
&(A_{\mu}(z,k)+\rho_{1}z\partial_{z}A_{\mu}(z,k))\vert_{z=1/\mu}=0 \label{bcA} \, .
\end{align}
\end{subequations}
The boundary conditions being homogeneous, it is obvious that they
introduce only IR data to the theory, and no dependence on the UV. In
other words, the different boundary conditions parametrize the way in
which conformal symmetry is spontaneously broken. Interestingly, we will actually see
that consistency and unitarity of the resulting field theory will
force us with a unique choice of boundary conditions. 

Through the holographic renormalization procedure, the resulting two-point functions are
\begin{subequations}
\begin{align}
&C_{2}(k^2)= C^{AdS}(k^2)+\frac{N^2}{6\pi^2}\frac{\rho_2 \frac{k}{\mu} K_{1}(\frac{k}{\mu})-K_2(\frac{k}{\mu})}{\rho_2\frac{k}{\mu} I_1(\frac{k}{\mu})+I_2(\frac{k}{\mu})} \\
&C_{3/2}(k^2)=C^{AdS}(k^2)+\frac{N^2}{6\pi^2}\frac{\rho_{3/2}\frac{k}{\mu}K_{1}(\frac{k}{\mu})-(1+\frac{\rho_{3/2}}{2})K_{2}(\frac{k}{\mu})}{(1+\frac{\rho_{3/2}}{2})I_{2}(\frac{k}{\mu})+\rho_{3/2}\frac{k}{\mu}I_{1}(\frac{k}{\mu})} \\
&C_{1R}(k^2)= C^{AdS}(k^2)+\frac{N^2}{6\pi^2}\frac{K_{1}(\frac{k}{\mu})-\rho_{1}\frac{k}{\mu}K_{0}(\frac{k}{\mu})}{I_{1}(\frac{k}{\mu})+\rho_{1}\frac{k}{\mu}I_{0}(\frac{k}{\mu})} \, ,
\end{align}
\end{subequations}
where $K_n$ and $I_n$ are (modified) Bessel functions. 

The trademark of the hard wall model is that correlation functions
approach their superconformal limit exponentially fast, at large
momentum. On the other hand, in the deep infrared the physics is determined by the choice of boundary conditions and in particular correlators can develop massless poles for specific choices of $\rho_s$. By expanding the above expression for $k^2/\mu^2\ll1$ we get
\begin{subequations}
\begin{align}
& C_{2}(k^2)\underset{\overset{{k^2\to0}}{}}{\simeq}\frac{N^2}{6\pi^2}\left(-\frac{16}{1+4\rho_{2}}\frac{\mu^4}{k^4}+\frac{16(1+6\rho_2(1+\rho_2))}{3(1+4\rho_2)^2}\frac{\mu^2}{k^2}+\dots\right) \\
& C_{3/2}(k^2)\underset{\overset{{k^2\to0}}{}}{\simeq}\frac{N^2}{6\pi^2}\left(-\frac{16(2+\rho_{3/2})}{(2+9\rho_{3/2})}\frac{\mu^4}{k^4}+\frac{16(4+\rho_{3/2}(28+37\rho_{3/2}))}{3(2+9\rho_{3/2})^2}\frac{\mu^2}{k^2}+\dots\right) \\
& C_{1R}(k^2)\underset{\overset{{k^2\to0}}{}}{\simeq}\frac{N^2}{6\pi^2}\left(\frac{2}{1+2\rho_{1}}\frac{\mu^2}{k^2}+\dots\right) \, .
\end{align}
\end{subequations}
All these expressions have poles for generic values of the boundary conditions. The appearance of double-poles in $C_2$ and $C_{3/2}$ is a sign of non-unitarity. Such double poles can (and have to) be cancelled by a specific choice of boundary conditions, i.e. $\rho_2\to\infty$ and  $\rho_{3/2}=-2$. This choice leaves us with form factors with only single poles, and  makes also $C_{2}(k^2)$ equal to $C_{3/2}(k^2)$. We then see that the only hard wall configuration which gives a dual QFT with a unitary spectrum has massless modes in both the stress-energy tensor and the supercurrent correlator, with positive residue. This shows that this configuration is mimicking a flow in which conformal symmetry is broken spontaneously. 

Since the theory is superconformal in the UV, supersymmetry cannot be
broken along the flow because having a non-zero vacuum energy would
contradict the operator identity $T=0$, which remains true when
conformal invariance is spontaneously broken. The $C_{1R}$ form factor
(which does not display double poles and hence does not have any
unitarity problem) is hence dictated by supersymmetry to be equal to $C_2$ and $C_{3/2}$, and this fixes the last parameter, $\rho_{1}=0$. This choice of boundary
condition for $A_\mu$ might be interpreted as the only one which corresponds 
to the correct superconformal R-current in the IR.

In summary, in the spontaneously broken conformal symmetry case we
have
\be
C_{2}(k^2)=C_{3/2}(k^2)=C_{1R}(k^2)=C^{AdS}(k^2)+\frac{N^2}{6\pi^2}\frac{K_{1}(\frac{k}{\mu})}{I_1(\frac{k}{\mu})}\underset{\overset{{k^2\to0}}{}}{\simeq}\frac{N^2}{6\pi^2}\frac{\mu^2}{k^2}+\dots \, .
\ee
The massless pole in the above form factors signals the presence of a
supermultiplet of massless particles in the dual field theory: these
are the dilaton for broken conformal symmetry \cite{Rattazzi:2000hs}, 
its superpartner the 
dilatino, and the R-axion, associated to the spontaneous breaking of the
superconformal R-symmetry. The presence of these strongly coupled
composite massless states nicely mirrors the same states that we found
in the weakly coupled model of section \ref{sectionmassless}. Note, however, the difference in the rest
of the spectrum. In the weakly coupled model one finds a massless
state and a continuum, after (possibly) a gap, while in the present
case it is easy to see, by continuing the Bessel functions to
negative values of $k^2$, that the spectrum is
composed exclusively of discrete states. 

\subsection{Explicitly broken conformal symmetry}

We now discuss the holographic version of a model with explicitly
broken conformal invariance but preserved supersymmetry. We expect 
$C_s$ form factors without massless
poles, and non-vanishing $F_s$ form factors.

We will consider the perturbation which breaks conformal invariance as given by a certain chiral operator $\mathcal{O}$ in the superpotential, dual to a hypermultiplet in the gravity theory. As anticipated, even if only a fully backreacted solution with a non-trivial profile for the hyperscalars can fully encode breaking of conformality, here we will take a short-cut. Our approximation consists in considering only the lowest order effects in the expansion parameter $m/\mu$, where $m$ is the scale of the perturbation, dual to the leading mode of the hyperscalar at the boundary, and $\mu$ is the scale of the IR wall. The operator $T_\mu^\mu$ and its supersymmetric partners have an explicit overall dependence on the scale $m$, reflecting the fact that they vanish in the limit $m\to 0$. In superfield language the relation reads
\be
X = \frac 43 \,(3-\Delta) \,m^{3-\Delta} \mathcal{O}\, ,
\ee
where $\Delta$ is the dimension of the operator ${\mathcal O}$, and $1 \leq \Delta <3$. It is clear, then, that to lowest order in $m/\mu$ the correlators of the trace operators are determined by those of $\mathcal{O}$ evaluated at $m=0$, i.e. in the conformal theory. This expansion corresponds, via holography, in an expansion in the profile of the hyperscalar dual to the coupling $m$. This argument then shows that the $F_s$ form factors can be obtained, to leading order, by simply fluctuating the hyperscalar dual to $\mathcal{O}$ in the background without any scalar profile, i.e. the hard wall. For a derivation of the precise relation between the correlators of $\mathcal{O}$ and the form factor $F_s$, see  appendix \ref{OWI} (the relations are derived there without reference to a small $m$ expansion, and therefore are valid independently from this limit). Note that, on the other hand, our crude approximation cannot capture the effect of the perturbation on the traceless part of two-point correlators. The 
dilaton, dilatino and axino should get a mass proportional to the scale $m$ of explicit breaking of conformal invariance, and correspondingly in the small $k^2$ limit the $C_s$ should take the gapped form $\sim (k^2 + m^2)^{-1}$. We expect this correction to be visible only working at higher order in the scalar profile. Already at the second order, however, the backreaction starts to be relevant, and therefore no calculation in the simple hard wall background can show this effect. 

Let us focus, for simplicity, on an operator with $\Delta=2$. The
relation between $X$ and ${\mathcal O}$ is in this case
\be
X=\frac43 \,m \,{\mathcal O}\, .
\ee
From appendix \ref{OWI}, we can read the relation between the $F_s$
form factors and the form factors of the operators in the chiral
multiplet ${\mathcal O}$
\be
F_2 =F_1= \frac 83 Z_F \quad , \quad  F_{3/2} = \frac 83 Z_\psi
\quad , \quad F_0 = \frac 83 Z_\phi  \, .
\ee
Implementing the holographic machinery we get
\begin{subequations}
\begin{align}
\label{Zpsi}
  Z_F (k^2)   &= Z^{ AdS}(k^2) +\frac{N^2}{4\pi^2}\frac{(1+\rho_1)K_{1}(\frac{k}{\mu})-\rho_{1}\frac{k}{\mu}K_{0}(\frac{k}{\mu})}{(1+\rho_1)I_{1}(\frac{k}{\mu})+\rho_{1}\frac{k}{\mu}I_{0}(\frac{k}{\mu})} 
\\
Z_\psi(k^2)&= Z^{ AdS}(k^2)+\frac{N^2}{4\pi^2}\frac{(1+\frac{3}{2} \rho_{1/2})K_{1}(\frac{k}{\mu})- \rho_{1/2}\frac{k}{\mu}K_{0}(\frac{k}{\mu})}{(1+\frac{3}{2} \rho_{1/2})I_{1}(\frac{k}{\mu})+\rho_{1/2}\frac{k}{\mu}I_{0}(\frac{k}{\mu})}
\\
 Z_\phi (k^2)&=Z^{ AdS}(k^2)+\frac{N^2}{4\pi^2}\frac{-(1+2\rho_{0})K_{0}(\frac{k}{\mu})+\rho_{0}\frac{k}{\mu} K_{1}(\frac{k}{\mu})}{(1+2\rho_{0})I_{0}(\frac{k}{\mu})+\rho_{0}\frac{k}{\mu}I_{1}(\frac{k}{\mu})}\, ,
\end{align}
\end{subequations}
where $Z^{ AdS}(k^2)$ is the usual conformal form factor containing
the $\log \Lambda^2/k^2$ term. Note that the non-trivial part of the
form factors is very similar to the ones computed in
\cite{Argurio:2012bi} for a vector supermultiplet, the dimensions
of the corresponding operators being the same. The parameters $\rho_1$, $\rho_{1/2}$
and $\rho_0$ are defined similarly as in \eqref{bch}--\eqref{bcA},
for the bulk fields of a hypermultiplet dual to 
${\mathcal O}$.

The only choice of parameters making all form factors equal and with
no massless poles is $\rho_0 = 0,\, \rho_1=-1,\,\rho_{1/2} =-\frac23$ which gives 
\begin{equation}
Z(k^2) = Z^{ AdS}(k^2)-\frac{N^2}{4\pi^2}\frac{K_{ 0}(\frac{k}{\mu})}{ I_{0}(\frac{k}{\mu})} \underset{\overset{{k^2\to0}}{}}{\simeq} \frac{N^2}{8\pi^2}\left(\log\frac{\Lambda^2}{\mu^2} -\frac{k^2}{2 \mu ^2} +O\left(k^4\right)\right)\,.
\end{equation}
Through Ward identities, this implies
that all $F_s$ form factors 
are non-vanishing, equal to one another, as expected, and gapped 
\begin{equation}
\label{Fbcc}
F_2(k^2) =F_{3/2}(k^2) =  F_1(k^2)=F_0(k^2) = \frac{N^2}{3\pi^2}\left(\log\frac{\Lambda^2}{k^2} -2 \frac{K_{ 0}(\frac{k}{\mu})}{ I_{0}(\frac{k}{\mu})} \right)\,.
\end{equation}

\subsection{Spontaneously broken supersymmetry}

We now consider the case of spontaneously broken supersymmetry. We remind that  
for this to be possible, conformal symmetry has to be
explicitly broken. In a supersymmetry breaking vacuum we expect a 
Goldstino and, specifically, a massless pole in the supercurrent
correlator. 
Using Ward identities as in the previous section, in particular eq. (\ref{traces-perturbation2}), 
this corresponds to a massless pole in the fermionic correlator $\langle \bar\psi_{O}(k) \psi_O(-k) \rangle $. In fact, for any choice of the parameter $\rho_{1/2}$ but the one discussed in the previous section, such a pole develops at low momenta
\begin{equation}
Z_\psi(k^2)\underset{\overset{{k^2\to0}}{}}{\simeq}\frac{N^2}{4\pi^2}
\frac{1+\frac{3}{2} \rho_{1/2}}{\frac12 + \frac74\rho_{1/2}}\frac{\mu^2}{k^2}+\dots
\end{equation}
Using 
\eqref{traces-perturbation2} we thus get, e.g. for $\rho_{1/2}=0$
\be
\langle \bar S_{\dot\alpha}(k)\,  S_\alpha(-k)\rangle =
{\sigma^\mu_{\alpha\dot\alpha}k_\mu}\frac{N^2}{\pi^2}
\frac{4 \, m^2\mu^2}{k^2}+\dots
\ee
This massless fermionic state, a composite state of the strongly
coupled gauge theory, is the Goldstino of spontaneously broken
supersymmetry. We have thus provided a holographic realization of the
Goldstino (albeit using the trick of the Ward identities) as the dual of
the lowest lying excitation of the fermionic operator in $\mathcal{ O}$. Note that here again we used the approximation of small $m/\mu$, and therefore the Goldstino propagator is expressed by the fermionic correlator evaluated in the conformal limit $m=0$. The scale of supersymmetry breaking $M$ can be read from the residue of the massless pole to be
\be
M = \sqrt{m\mu}\, .
\ee
This approximate formula nicely reflects that the effect responsible for the breaking of supersymmetry are the boundary conditions at the IR wall ($M=0$ when $\mu=0$) and also that conformal symmetry must be explicitly broken to have a non-supersymmetric vacuum ($M=0$ when $m=0$).

In order to go beyond the lowest order in $m/\mu$ and find a massless pole in the supercurrent correlator directly, we would
need a backreacted space-time with scalar profiles that break
supersymmetry by sub-leading modes (i.e. corresponding to the VEV of some F-term in the field theory). The latter would also be the only approach that would
give us a non-vanishing one-point function $\langle T_{\mu\nu}\rangle$.

As a final remark, let us notice that there is in fact a special
choice of parameters which, while keeping the massless pole in the fermionic correlator, makes all form factors equal, namely  $\rho_0 = -\frac12,\, \rho_1=0,\,\rho_{1/2} =0$. This corresponds to a common $Z$ form factor
\begin{equation}
Z (k^2)= \frac{N^2}{8\pi^2}\left(\log\frac{\Lambda^2}{k^2} + 2
\frac{K_{ 1}(\frac{k}{\mu})}{ I_{1}(\frac{k}{\mu})}\right) \underset{\overset{{k^2\to0}}{}}{\simeq}\frac{N^2}{2\pi^2}
  \frac{ \mu ^2}{k^2}\, .
\end{equation} 
This gives $1/k^2$ poles at low momenta for all real correlators of
operators in the FZ multiplet. While such result might be interpreted
as a supersymmetric vacuum with a massless chiral superfield in an
otherwise gapped spectrum, the most natural interpretation is in fact
that the apparent spectrum degeneracy is just an accident of the
specific model. This is  reminiscent of  a Polonyi model
which, while breaking supersymmetry,  has a massless supersymmetric spectrum as
the Goldstino is matched with a pseudomodulus and an R-axion.

\section{Summary and outlook}

In this paper we have studied two-point functions of operators belonging to the supercurrent multiplet(s) of ${\mathcal N}=1$ supersymmetric field theories, parametrizing the correlators in terms of momentum dependent form factors. We have discussed explicit field theory examples, both weakly and strongly coupled,  in different dynamical phases: we considered superconformal theories, both in symmetry preserving vacua and in vacua with spontaneously broken conformal symmetry, as well as non-conformal ones, both in supersymmetry preserving and breaking vacua. 

In the holographic context we focused on pure AdS and hard wall backgrounds. While the former case represents vacua preserving superconformal symmetry, the latter describes vacua where conformal symmetry is spontaneously broken, and massless poles associated to the corresponding Goldstone modes appear. In order to describe non-conformal theories holographically, one should consider less trivial backgrounds, in which additional hypermultiplets, dual to a superpotential perturbation, have non-trivial profiles, and as such backreact on the metric, deforming the AdS-ness of the background. Still, we have shown that working at the leading order in the perturbation, one can get non-trivial traceful contributions to the correlators by evaluating hypermultiplet two-point functions in the unperturbed, purely hard wall, background. This is just the leading contribution to the $F_s$ form factors, of course, but the only one the hard wall can capture. Finally, by considering non-supersymmetric IR  boundary 
conditions for the hypermultiplet, we were also able to realize a holographic toy-model of spontaneous supersymmetry breaking, and to show that the supercurrent correlator has the expected massless pole corresponding to the Goldstino.

The holographic model we have used in this work, despite the virtue of being flexible and easily calculable, is not obtained as a solution of the supergravity equations of motion. One obvious future direction would be to work at the level of a consistent ${\mathcal N}=2$ truncation of ${\mathcal N}=8$ gauged five-dimensional supergravity, and consider backreacted backgrounds, such as (non-supersymmetric deformations of) those discussed in \cite{Girardello:1998pd,Freedman:1999gp,Girardello:1999bd,Ceresole:2001wi,Bertolini:2013vka}. In such models, one would be able to compute holographically $C_s$ and $F_s$ form factors for non-conformal theories,  to all orders in the relevant perturbation.

Our approach could also be useful to analyze supersymmetry breaking models in the context of string theory, and possibly consider backgrounds which are not asymptotically AdS, as for example the one discussed in \cite{Kachru:2002gs,DeWolfe:2008zy,Bena:2009xk,Dymarsky:2011pm}. Indeed, two-point correlators can be effectively used as a probe of the dynamics which breaks supersymmetry, for instance by discriminating an explicit breaking from a spontaneous one. To this aim, a discerning result would be to obtain, via holography, the massless pole associated to the Goldstino.

\section*{Acknowledgments}

We are grateful to G. Barnich, F. Bastianelli, Z. Komargodski, R. Rahman and A. Schwimmer for useful discussions. The research of R.A. and D.R. is supported in part by IISN-Belgium (conventions 4.4511.06, 4.4505.86 and 4.4514.08), by the ``Communaut\'e
Fran\c{c}aise de Belgique" through the ARC program and by a ``Mandat d'Impulsion Scientifique" of the F.R.S.-FNRS. R.A. is a Senior Research Associate of the Fonds de la Recherche Scientifique--F.N.R.S. (Belgium). M.B., L.D.P. and F.P. acknowledge partial financial support by the MIUR-PRIN contract 2009-KHZKRX. L.D.P. is supported by the ERC STG grant number 335182, by the Israel Science Foundation under grant number 884/11. L.D.P. would also like to thank the United States-Israel Binational Science Foundation (BSF) for support under grant number 2010/629. In addition, the research of L.D.P. is supported by the I-CORE Program of the Planning and Budgeting Committee and by the Israel Science Foundation under grant number 1937/12.   Any opinions, findings, and conclusions or recommendations expressed in this material are those of the authors and do not necessarily reflect the views of the funding agencies.
~

\appendix

\section{Supersymmetry transformations}
\label{AppTrans}
In this appendix we collect the supersymmetry transformations between operators belonging to the FZ and ${\mathcal R}$ multiplets. The supersymmetric variation of the operators in the FZ multiplet can be obtained from the component expressions \eqref{J-comp} and \eqref{X-comp}. The result can be summarized as follows
\begin{subequations}\label{susyvar}
\begin{align}
   \delta x & = \frac 2 3 \,\epsilon S\,, \label{deltax}
\\ \delta j_\mu & = \epsilon\left( S_\mu - \frac 1 3 \sigma_\mu \bar S \right) + \bar\epsilon \left( \bar S_\mu + \frac 1 3 \bar\sigma_\mu S \right)\,,
\\ \delta S_{\mu\alpha} & = 2i\,(\sigma_{\mu\nu}\epsilon)_\alpha\,\partial^\nu x^* + (\sigma^\nu\bar\epsilon)_\alpha\left( 2\timn +i\partial_\nu j_\mu-i\nmn \partial_\rho j^\rho + \frac1 2 \varepsilon_{\mu\nu\rho\lambda} \partial^\rho j^\lambda \right)\,,
\\ \delta\timn & = -i\, \epsilon\,  \sigma_{\rho(\mu} \partial^\rho S_{\nu)}+ i\, \bar\epsilon\, \bar\sigma_{\rho(\mu} \partial^\rho \bar S_{\nu)}\,,
\end{align}
\end{subequations}
where the indices between round brackets are symmetrized with the combinatorial factor. We also list, below, the supersymmetry transformation for the trace operators of the FZ multiplet and the divergence of the current
\begin{subequations}
\begin{align}
   \delta S & = \epsilon \left( 2\, T + 3\, i\, \partial_\mu j^\mu \right) + 3\, i\, \sigma^\mu\bar\epsilon\, \partial_\mu x \,,
\\ \delta T & = \frac i 2 \epsilon\, \sigma^\mu\partial_\mu S + \frac i 2 \bar\epsilon\, \bar\sigma^\mu\partial_\mu\bar S \,,
\\ \delta \left(\partial_\mu j^\mu  \right) & = -\frac 1 3 \epsilon\, \sigma^\mu\partial_\mu\bar S + \frac 1 3 \bar\epsilon\, \bar\sigma^\mu\partial_\mu\bar S \,.
\end{align}
\end{subequations}
Notice that these last three variations plus \eqref{deltax} close the algebra on their own (indeed, they make up the chiral multiplet $X$ defined in \eqref{X-comp}). This is also consistent with the superconformal case, where these four operators can all be consistently set to zero.

The supersymmetry transformations of the fields belonging to the ${\mathcal R}$ multiplet read
\begin{subequations}\label{susyvarR}
\begin{align}
\delta j_\mu & = \epsilon S_\mu + \bar\epsilon \bar S_\mu \,,
\\ \delta S_\mu & =  \sigma^\nu \bar\epsilon \left( i \partial_\nu j_\mu + 2\timn + \frac1 2 \varepsilon_{\mu\nu\rho\sigma} (\partial^\rho j^\sigma + C^{\rho\sigma})\right)\,,
\\ \delta\timn & = -i\, \epsilon\,  \sigma_{\rho(\mu} \partial^\rho S_{\nu)}+ i\, \bar\epsilon\, \bar\sigma_{\rho(\mu} \partial^\rho \bar S_{\nu)}\,,
 \\  \delta C_{\mu\nu} & = \epsilon \, \sigma_{[\mu}\partial_{\nu]}\,\bar{S}-\bar{\epsilon}\, \bar{\sigma}_{[\mu}\partial_{\nu]} \, S \label{deltaC}\ .
\end{align}
\end{subequations}

\section{Perturbation of the fixed point and non-conformal form factors}
\label{OWI}

In the general parametrization of correlators in terms of form factors of section \ref{formfactors}, it has been stressed that some of them are generated only when conformal symmetry is explicitly broken. In this appendix we will show that non-conformal form factors are in fact determined by  correlators of the operator which perturbs the fixed point and starts the RG flow. We will do this for the FZ multiplet, and briefly comment on the analogous relations for the ${\mathcal R}$ multiplet. The Lagrangian is that of a SCFT, perturbed by a relevant operator. As shown in \cite{Green:2010da}, the only possible relevant deformation is given by a superpotential, namely by a chiral operator ${\mathcal O}$ of dimension $\Delta$ with $1\leq \Delta < 3$
\begin{align}
& \bar D_{\dot \alpha}\mathcal{O} = 0\,,\quad {\mathcal O} = \phi_O + \sqrt{2} \theta \psi_O + \theta^2 F_O + \dots  \\
& {\mathcal L} = {\mathcal L}_{SCFT} + m^{3-\Delta} F_O + c.c.\,.
\end{align}
We can parametrize the real two-point functions of $\mathcal{O}$ in terms of the following real form factors
\begin{subequations}
\begin{align}
&\langle \phi_O^*(k) \phi_O(-k)\rangle = m^{2\Delta - 4} \, Z_\phi  \\ 
&\langle {\bar{\psi}_O}_{\dot{\alpha}}(k) {\psi_O}_\alpha(-k) \rangle = m^{2\Delta - 4} \, \sigma^{\mu}_{\alpha\dot{\alpha}}k_\mu Z_\psi \\ 
&\langle F_O^*(k) F_O(-k)\rangle = - m^{2\Delta - 4} \, k^2 Z_F\,,  
\end{align}
\end{subequations}
and the following complex form factors
\begin{subequations}
\begin{align}
 & \langle \phi_O(k) \phi_O(-k)\rangle = m^{2\Delta -4} \, Y_\phi \\ 
 &\langle{\psi_O}_\alpha(k){\psi_O}_\beta(-k)\rangle = m^{2\Delta - 3} \epsilon_{\alpha\beta} \, Y_\psi  \\ 
 & \langle F_O(k) F_O(-k) \rangle = m^{2\Delta -4} \, k^2 Y_F  \\ 
 & \langle \phi_O(k)F_O(-k)\rangle = m^{2\Delta -3} \, Y_{\phi\,F} \\ 
&\langle \phi_O^*(k) F_O(-k)\rangle = m^{2\Delta -3} \,{\tilde{Y}}_{\phi\,F}  \, .
 \end{align}
 \end{subequations}
 In a vacuum which preserves supersymmetry, the following relations hold
\be
Z_\phi = Z_\psi = Z_F \,, \quad Y_\psi = Y_{\phi\,F} \,, \quad Y_\phi = Y_F = \tilde{Y}_{\phi\,F} = 0 \,. 
\ee
The relation between the chiral superfield $X$ of the FZ multiplet and the operator $\mathcal{O}$ reads
\be
\label{XO}
X = \frac 43 (3-\Delta)\,m^{3-\Delta} \mathcal{O}\,,
\ee
which implies the following relations between the correlators (up to possible contact terms, because the relation is only valid on-shell)
 \begin{subequations}
 \begin{align}
\langle T(k) T(-k) \rangle =& \, 2 (3-\Delta)^2 \, m^{6-2\Delta} \, ( \mathrm{Re}\langle F_O(k) F_O(-k) \rangle + \langle F_O^*(k) F_O(-k) \rangle  )  \label{traces-perturbation1}\\
\langle \bar{S}_{\dot{\alpha}}(k) S_\beta(-k) \rangle = & \, 8 (3- \Delta)^2 \, m^{6-2\Delta} \, \langle {\bar{\psi}_O}_{\dot{\alpha}}(k) {\psi_O}_\beta(-k) \rangle \label{traces-perturbation2}
 \\
\langle \partial j (k) \partial j (-k) \rangle = & \, \frac 89 (3- \Delta )^2 \, m^{6-2\Delta} \, (- \mathrm{Re}\langle F_O(k) F_O(-k) \rangle + \langle F_O^*(k) F_O(-k) \rangle) 
\\
\langle  x^* (k) x(-k) \rangle = & \frac {16}9 (3- \Delta)^2 m^{6-2\Delta} \langle \phi_O^*(k) \phi_O(-k) \rangle\, .\label{traces-perturbation4}                                       
\end{align}
\end{subequations}
Comparing with eqs.~(\ref{<TTtr>})-(\ref{xxdagger2}), one gets for the FZ form factors 
\begin{subequations}
\begin{align}
\label{mWI}
F_2 &= \frac 83 \left(3-\Delta\right)^2 \left(Z_F - \mathrm{Re} Y_F \right)\\
F_{ 3/2} + \frac{32}{3}  \frac{M^4}{m^2k^2} & = \frac 83 (3-\Delta)^2 Z_\psi \label{relFO32}\\
F_1 &= \frac 83 (3-\Delta)^2 \left( Z_F + \mathrm{Re} Y_F \right)  \\
F_0 & = \frac 83 (3-\Delta)^2  Z_\phi. 
\end{align}
 \end{subequations}
In eq.~(\ref{relFO32}) the  additional term displaying the expected massless pole associated to the Goldstino is present, see eq.~(\ref{<SSbar2>}).

Let us also mention the case of the ${\mathcal R}$ multiplet. In this case, the operator giving the superpotential perturbation is related on-shell to a real superfield $\mathcal{O}_R$
\begin{equation}
\mathcal{O} = \bar{D}^2 \mathcal{O}_R.
\end{equation}

The relation with the operator $\chi_\alpha$ that contains the trace is
\be
\chi_\alpha = - 4 \, (3 - \Delta) \,  m^{3-\Delta}  \, \bar{D}^2 \, D_\alpha \, \mathcal{O}_R
\ee
and the non-conformal form factors in this case can be expressed in terms of those of the operator $\mathcal{O}_R$.

\section{Traceless Form Factors in Holography}
\label{appholo}
In this section we give some details about the holographic computation of the
traceless form factors $C_{s}(k^2)$. We will consider a quadratic action describing free fluctuations of the supergravity bulk multiplet $\{h_{MN},\,\psi_{M},\, A_{M}\}$  over an $\text{AdS}_5$ background. This is enough for computing two-point functions of the stress-energy tensor $T_{\mu\nu}$,  supercurrent $S_{\mu}$ and superconformal R-symmetry current $j_{\mu}$ in the QFT dual to either pure $\text{AdS}_5$ or $\text{AdS}_5$ with a hard wall. The supergravity action reads
\be
S=\frac{N^2}{4\pi^2}\int \text{d}^5x\sqrt{G}\Big(-\frac12 R-6
+\bar{\psi}_M(\Gamma^{MNP}{D}_N-\frac32\Gamma^{MP})\psi_P+
\frac14 G^{MP}G^{NQ}F_{MN}F_{PQ}\Big)\, , \label{bulkactionapp}
\ee
The overall constant is fixed in terms of the $AdS_5\times S^5$ ten-dimensional solution, $\frac{1}{8\pi
G_\mathrm{N}}=\frac{N^2}{4\pi^2}$ with $L_\mathrm{AdS}=\alpha'=1$,
and we use indices such that $x^M=(z,x^\mu)$. 
The $\text{AdS}_5$ background metric is
\begin{equation}
 ds^2 =\frac{1}{z^2}\eta_{MN}dx^Mdx^N= \frac{1}{z^2}\left( dz^2 + \eta_{\mu\nu}dx^\mu dx^\nu \right)\,
\end{equation}
and the graviton field $h_{MN}$ is defined as the fluctuation around $\eta_{MN}$. As usual we can exploit bulk gauge freedom and consider fluctuations in the axial gauge $A_{z}=h_{M z}=\psi_{z}=~0$. Inspection of the $\text{AdS}_5$ equations of motion reveals that the transverse-traceless part of the bulk fields decouple from the rest and satisfy homogeneous ordinary differential equations which after Fourier-transforming from $x_\mu$ to $k_\mu$ read%
\footnote{Notice that, analogously to the case of the spinor previously
discussed in \cite{Argurio:2012cd,Argurio:2012bi}, we have traded the first order equation of motion for a Dirac field with a second order equation of motion for one of its Weyl components plus a first order constraint for the other Weyl component, choosing
\be
\psi_{\mu}= \left(\begin{array}{c}\xi_{\mu} \\ \bar \chi_{\mu}\end{array}\right)\ .
\ee
}
\begin{align}
&\ \ \ (z^2\partial_{z}^2-3z\partial_{z}-z^2k^2)h^{tt}_{\mu\nu}(z,k)=0 \label{eqAdS2}\\
&\begin{cases}
(z^{2}\partial_{z}^2-4z\partial_{z}-z^{2}k^2+\frac{9}{4})\xi^{tt}_\mu(z,k)=0 \label{eqAdS3/2}\\
z\sigma^{\nu}k_{\nu}\bar{\chi}^{tt}_{\mu}(z,k)=(-z\partial_{z}+\frac{1}{2})\xi^{tt}_{\mu}(z,k)\, 
\end{cases}\\
&\ \ \ (z^2\partial_{z}^2-z\partial_{z}-z^2k^2) A^t_{\mu}(z,k)=0 \label{eqAdS1}\ ,
\end{align}
The constraints $h^{tt\mu}_{\phantom{tt}\mu}=\partial^{\mu}h^{tt}_{\mu\nu}=0$, $\gamma^{\mu}\psi^{tt}_{\mu}=\partial^{\mu}\psi^{tt}_{\mu}=0$ and  $\partial^{\mu}A^t_{\mu}=0$ can be implemented by means of appropriate projectors acting on the complete fluctuations 
\begin{equation}\label{projector}
h_{\mu\nu}^{tt}=-\frac{1}{3k^4}X_{\mu\nu}^{\rho\sigma}h_{\rho\sigma}\ ,\qquad \bar{\xi}^{tt}_\mu=\frac{2\bar{\sigma}^{\rho}k_\rho}{3k^4}Y_{\mu}^{\nu}\bar{\xi}_\nu\ , \qquad  A^t_{\mu}=\frac{P_{\mu}^{\nu}}{k^2}A_{\nu}\ ,
\end{equation}
where $X_{\mu\nu}^{\rho\sigma}$ and $Y_{\mu}^{\nu}$ are defined in \eqref{X} and \eqref{Y} respectively and $P_{\mu}^{\nu}$ is the usual transverse projector. 
Since we are only interested in computing transverse-traceless form factors we can focus on the $tt$ part of the bulk field and disregard the rest of the equations of motion. For ease of notation we will omit the $tt$ superscript in the rest of the discussion.

Solutions to the above differential equations behave near $z=0$ as
\begin{align}
&h_{\mu\nu}(z,k) \underset{\overset{\mathrm{z\to0}}{}}{\simeq}h_{0\,\mu\nu}(k)+z^2\,h_{2\,\mu\nu} (k) +z^4\,\log(z\Lambda)\,h_{4\,\mu\nu} (k)+z^4\,\tilde{h}_{4\,\mu\nu} (k)+\mathcal{O}(z^6) \label{boundary2}\\
& \begin{cases} 
\xi_{\mu}(z,k)\underset{\overset{\mathrm{z\to0}}{}}{\simeq}z^{1/2}\left(\xi_{0\,\mu}(k)+z^{2}\,\xi_{2\,\mu}(k)+z^4\,\log(z\Lambda)\,\xi_{4\,\mu}(k)+z^{4}\,\tilde{\xi}_{4\,\mu}(k)+\mathcal{O}(z^6)\right)\\
\bar{\chi}_{\mu}(z,k)\underset{\overset{\mathrm{z\to0}}{}}{\simeq}z^{3/2}\left(\bar{\chi}_{0\,\mu}(k)+z^2\,\log(z\Lambda)\,\bar{\chi}_{2\,\mu}(k)+z^2\,{\bar{\tilde{\chi}}}_{2\,\mu}(k)+\mathcal{O}(z^4)\right)\ 
\end{cases}\\
&A_{\mu}=a_{0\,\mu}(k)+z^2\,\log(z\Lambda)\,a_{2\,\mu}(k)+z^2\,\tilde{a}_{2\,\mu}(k)+\mathcal{O}(z^4)
\label{boundary1}\ .
\end{align}
The coefficients of the near-boundary expansion satisfy the following relations
\begin{align}
&h_{2\,\mu\nu}(k)=-\frac{k^2}{4}h_{0\,\mu\nu}(k)\ ,\ h_{4\,\mu\nu}(k)=-\frac{k^4}{16}h_{0\,\mu\nu}(k)\ , \\
&\xi_{2\,\mu}=-\frac{k^2}{4}\xi_{0\,\mu}\ ,\ \xi_{4\,\mu}=-\frac{k^4}{16}\xi_{0\,\mu}\ ,\notag\\
&\bar{\chi}_{0\,\mu}=-\frac{1}{2}\bar{\sigma}^{\nu}k_{\nu}\,\xi_{0\,\mu}\ ,\ \bar{\chi}_{2\,\mu}=-\frac{k^2}{4}\bar{\sigma}^{\nu}k_{\nu}\xi_{0\,\mu}\ ,\ \bar{\tilde{\chi}}_{2\,\mu}=- 4\frac{\bar{\sigma}^{\nu}k_{\nu}}{k^2}\,\tilde{\xi}_{4\,\mu}+\frac{k^2}{16}\bar{\sigma}^{\nu}k_{\nu}\xi_{0\,\mu}\ , \\
&a_{2\,\mu}(k)=\frac{k^2}{2}a_{0\,\mu}(k).
\end{align}
The leading terms $\lbrace  h_0^{\mu\nu}(k),
\xi_0^\mu(k), a_0^\mu(k)\rbrace$ are identified as the sources of the
corresponding boundary operators $\lbrace  T_{\mu\nu}(k),
S_{\mu}(k), j_\mu(k)\rbrace$. Note that the scaling behavior at the boundary,
which depends on the mass of the fluctuating field in
$\text{AdS}_5$, is the correct one to get a multiplet of operators of
dimension $\lbrace4,7/2, 3\rbrace$ respectively. Also, having chosen a
positive sign for the mass for the gravitino field, the leading terms
at the boundary has positive chirality. The undetermined subleading
terms $\lbrace \tilde{h}_{4\,\mu\nu}(k),
\tilde{\xi}_{4\,\mu}(k),\tilde{a}_{2\,\mu}(k)\rbrace$ are associated to the one-point
functions of the boundary operators, and their functional dependence on the sources will be determined imposing boundary conditions in the bulk. 

The on-shell boundary action at the regularizing surface $z=\epsilon$ is 
\begin{equation}
S_{reg}= \frac{N^2}{4\pi^2}\int_{z=\epsilon}\frac{d^4k}{(2\pi)^4}\left[\frac{1}{4z^3}h^{\mu\nu}h'_{\mu\nu}-\frac{3}{2z^4}h^{\mu\nu}h_{\mu\nu}+6+\frac{1}{2z^4}(\xi^\mu\chi_\mu+\bar{\chi}^{\mu}\bar{\xi}_{\mu})+\frac{1}{2z}A^{\mu}\partial_{z}A_{\mu}\right]\ .
\end{equation}  
Here the 4d space-time indices are raised and lowered using the flat metric and we have added all the boundary terms that are needed to have a well defined variational principle \cite{Volovich:1998tj,Rashkov:1999ji,Amsel:2009rr}.

The above action can be made finite by adding appropriate 4d-covariant counterterms at the regularizing surface \cite{deHaro:2000xn, Bianchi:2001de, Bianchi:2001kw}
\begin{align}
S_{ct}=\frac{N^2}{4\pi^2}\int_{z=\epsilon}\frac{d^4k}{(2\pi)^4}\sqrt{\gamma}&\left[ 6-R[\gamma]+(\log(\epsilon\Lambda)+\alpha_2)\frac{R^{\mu\nu}R_{\mu\nu}[\gamma]}{4} -\frac{i}{2}\bar{\psi}^{\mu}\gamma^{\nu}k_{\nu}\psi_{\mu}\right.\notag\\
&\left.+\frac{i}{4}(\log(\epsilon\Lambda)+\alpha_{3/2})\bar\psi^{\mu}k^2\gamma^{\nu}k_{\nu}\psi_{\mu}+\frac{1}{4}(\log(\epsilon\Lambda)+\alpha_1)F^{\mu\nu}F_{\mu\nu}\right]\ ,
\end{align}
where we have defined the metric at the regularizing surface as $\gamma_{\mu\nu}=\frac{1}{\epsilon^2}\left(\eta_{\mu\nu}+h_{\mu\nu}\right)$ (meaning that 4d space-time indices are raised and lowered using $\gamma_{\mu\nu}$) and the action should be intended up to quadratic order in the fields. Notice that the counterterms are defined up to possible finite contributions $\alpha_s$ which distinguish between different renormalization schemes. The resulting renormalized action $S_{ren}=S_{reg}+S_{ct}$ can be expressed purely in terms of the leading and the subleading modes of the fluctuations
\begin{align}
S_{ren}=\frac{N^2}{4\pi^2}\int\frac{d^4k}{(2\pi)^4}&\left[\frac{1}{6k^4}h_{0\mu\nu}X_{\mu\nu}^{\rho\sigma}\tilde{h}_{4\,\rho\sigma}+\frac{8}{3k^4}\xi_0 ^{\mu}Y_{\mu}^{\nu}\bar{\tilde{\xi}}_{4\nu}-\frac{1}{k^2}a_{0\,\mu}P_{\mu}^{\nu}\, \tilde{a}_{2\,\nu}\right.\nonumber \\
&\left. + \,\text{terms quadratic in the sources}\right]\,.
\end{align}
where we reinstated the appropriate projectors using \eqref{projector}. The operators of the boundary theory are defined through the AdS/CFT correspondence as the composite operators sourced by the leading modes of each bulk fluctuation
\begin{equation}
S_{int}[h^{\mu\nu}_{0},\xi^{\mu}_{0},\bar{\xi}^{\mu}_{0},a^{\mu}_{0}]= \int\frac{d^4k}{(2\pi)^4}\left[\frac{1}{2}h^{\mu\nu}_{0}T_{\mu\nu}+\frac 12(\xi^{\mu}_{0}S_{\mu}+\bar{\xi}^{\mu}_0\bar{S}_{\mu})+\sqrt{\frac{3}{2}}
  a^{\mu}_{0}j_{\mu}\right]\ .
\end{equation}
where the relative coefficients between the different terms are fixed by supersymmetry \cite{Komargodski:2010rb}. 

The corresponding two-point functions are then obtained differentiating twice the renormalized action with respect to the sources 
\begin{align}
\langle T_{\mu\nu}^{tt}(k)T_{\rho\sigma}^{tt}(-k)\rangle&=\!-4\frac{\delta^2 S_{ren}}{\delta h^{\mu\nu}_0\delta h^{\rho\sigma}_0}=\!-\frac{N^2}{6\pi^2}\left[X_{\mu\nu}^{\gamma\delta}\left(\frac{\delta\tilde{h}_{4\gamma\delta}}{k^4\delta h_{0}^{\rho\sigma}}+\alpha'_{2}\eta_{\sigma\rho}\eta_{\sigma\delta}\right)+(\mu,\nu)\leftrightarrow(\rho,\sigma)\right]\ \notag \\ 
\langle S_{\mu}(k)\bar{S}_{\nu}(-k)\rangle&=\!-4\frac{\delta^2 S_{ren}}{\delta\xi_{0}^{\mu}\delta\bar{\xi}_{0}^{\nu}}=-\frac{8 N^2}{3\pi^2}\left[Y_{\mu}^{\rho}\left(\frac{\delta{\bar{\tilde{\xi}}}_{4\rho}}{k^4\delta\bar{\xi}_{0}^{\nu}}+\alpha'_{3/2}\eta_{\rho\nu}\right)\right] \\
\langle j_\mu(k)j_\nu(k)\rangle&=\!-\frac{2}{3}\frac{\delta^2 S_{ren}}{\delta a^{\mu}_{0}\delta a^{\nu}_{0}}=\frac{N^2}{6\pi^2}\left[P_{\mu}^{\rho}\left(\frac{\delta\tilde{a}_{2\rho}}{k^2\delta a^{\nu}_{0}}+\alpha'_{1}\eta_{\rho\nu}\right)+\mu\leftrightarrow\nu\right]\ ,\notag
\end{align}
where the constants $\alpha'_s$ can be written in terms of the finite counterterms coefficients $\alpha_s$.
In presenting our results we choose a subtraction
scheme in which all the finite contributions deviating from pure logarithmic behavior in the superconformal case are reabsorbed by
finite counterterms. 

\let\oldbibitem=\bibitem\renewcommand{\bibitem}{\filbreak\oldbibitem}
\bibliographystyle{JHEP}
\fussy
\bibliography{Biblio}

\end{document}